\documentclass{aastex}
\usepackage{graphicx,epsfig,emulateapj5}

\newcommand{\etal}{{et al.~}}

\newcommand{\bq}{\begin{equation}}
\newcommand{\eq}{\end{equation}}
\newcommand{\msun}{\mbox{M$_{\odot}$}}

\def\gtsim{\lower.5ex\hbox{$\buildrel > \over\sim$}}
\def\ltsim{\lower.5ex\hbox{$\buildrel < \over\sim$}}
\def\arcsec{^{\prime\prime}}

\def\sun{\mbox{$_\odot$}}
\def\apjl{ApJL}
\def\apj{ApJ}
\def\apjs{ApJS}
\def\mnras{MNRAS}
\def\araa{ARAA}
\def\aj{AJ}
\def\aap{A\&A}
\def\aaps{A\&A Suppl.}
\def\nat{Nature}

\begin{document}
\title
{The HST/ACS Coma Cluster Survey. VIII. Barred Disk Galaxies in the Core of the Coma Cluster\altaffilmark{1}}

\author{Irina Marinova\altaffilmark{2}, 
Shardha Jogee\altaffilmark{2}, 
Tim Weinzirl\altaffilmark{2},
Peter Erwin\altaffilmark{3,4},
Neil Trentham\altaffilmark{5},
Henry C.~Ferguson\altaffilmark{6},
Derek Hammer\altaffilmark{7}, 
Mark den Brok \altaffilmark{8},
Alister W.~Graham\altaffilmark{9}, 
David Carter\altaffilmark{10},
Marc Balcells\altaffilmark{11},
Paul Goudfrooij\altaffilmark{6},
Rafael Guzm\'an\altaffilmark{12},
Carlos Hoyos\altaffilmark{12},
Bahram Mobasher\altaffilmark{13},
Mustapha Mouhcine\altaffilmark{10},
Reynier F.~Peletier\altaffilmark{8},
Eric W.~Peng\altaffilmark{14,15},
and Gijs Verdoes Kleijn\altaffilmark{8},
}   
\authoremail{marinova@astro.as.utexas.edu, sj@astro.as.utexas.edu}
\altaffiltext{1}{Based on observations with the NASA/ESA Hubble Space
Telescope obtained at the Space Telescope Science Institute,
which is operated by the Association of Universities for Research
in Astronomy, Inc., under NASA contract NAS 5-26555. These
observations are associated with program GO10861.}
\altaffiltext{2}{Department of Astronomy, University of Texas at
Austin, Austin, TX}   
\altaffiltext{3}{Max-Planck-Insitut f\"{u}r extraterrestrische Physik, Giessenbachstrasse, 85748 Garching, Germany}
\altaffiltext{4}{Universit\"{a}ts-Sternwarte M\"{u}nchen, Scheinerstrasse 1, 81679 M\"{u}nchen, Germany}
\altaffiltext{5}{Institute of Astronomy, Madingley Road, Cambridge CB3 0HA}
\altaffiltext{6}{Space Telescope Science Institute, 3700 San Martin Drive, 
Baltimore, MD 21218, USA.}
\altaffiltext{7}{Department of Physics and Astronomy, Johns Hopkins, Baltimore, MD 21218, USA.}
\altaffiltext{8}{Kapteyn Astronomical Institute, University of Groningen, Groningen, The Netherlands.}
\altaffiltext{9}{Centre for Astrophysics and Supercomputing, Swinburne 
University, Hawthorn, Australia.}
\altaffiltext{10}{Astrophysics Research Institute, Liverpool John Moores 
University,  Birkenhead, UK.}
\altaffiltext{11}{Instituto de Astrof\'isica de Canarias, 
38200 La Laguna, Tenerife, Spain.}
\altaffiltext{12}{Department of Astronomy, University of Florida, Gainesville, FL 32611, USA.}
\altaffiltext{13}{Department of Physics and Astronomy, University of
  California, Riverside, CA 92521, USA}
\altaffiltext{14}{Department of Astronomy, Peking University, Beijing 100871, China}
\altaffiltext{15}{Kavli Institute for Astronomy and Astrophysics, Peking University, Beijing 100871, China}

\begin{abstract}
We use high resolution  ($\sim0.1\arcsec$)
F814W ACS images from the \textit{Hubble Space Telescope}  ACS
Treasury survey of the Coma cluster at $z\sim$~0.02 to study bars in 
massive disk galaxies (S0s), as well as low-mass dwarf galaxies in
the  core of the Coma cluster,  the densest environment in the nearby 
Universe. Our study  helps to  constrain the evolution of bars and
disks in dense environments and provides a  comparison point for  studies 
in lower density environments and at higher redshifts. Our results are:
(1)~We characterize the fraction and properties of bars in 
a sample of 32 bright ($M_{\rm V}\lesssim -18$, $M_{*} > 10^{9.5}
M_{\odot}$) S0 galaxies, which dominate the population of massive 
disk galaxies in the Coma core.  We find that 
the  measurement of a bar fraction among S0 galaxies must be handled
with special care due to the difficulty in separating
unbarred S0s from ellipticals, and the potential dilution  of the bar
signature by light from a relatively large, bright bulge.  The results
depend sensitively on the method used: the bar fraction for bright S0s in 
the Coma  core is 50$\pm$~11\%, 65$\pm$~11\%, and 60$\pm$~11\% 
based on three methods of bar detection, namely  strict ellipse fit
criteria, relaxed ellipse fit criteria, and visual classification.
(2)~We compare the S0 bar fraction across different environments (the Coma core, Abell 901/902, and Virgo)
adopting the critical step of using matched samples and matched 
methods in order to ensure robust comparisons. 
We find  that the  bar fraction  among bright S0 galaxies does not 
show a statistically significant variation  (within the error bars 
of $\pm$11\%)  across environments which span two orders
of magnitude in galaxy number density 
($n\sim$~300--10,000~gal/Mpc$^{3}$), and include 
rich and poor clusters, such as  the core of Coma,  the Abell 901/902  
cluster, and Virgo.
We speculate that the bar fraction  among S0s is not significantly 
enhanced in rich  clusters compared to low density environments
due to two reasons. Firstly, S0s in rich clusters  are less prone 
to bar instabilities as they are dynamically heated by harassment 
and are gas poor as a result of ram pressure  stripping and 
accelerated star formation. Secondly,  high-speed encounters in 
rich clusters may be less effective than slow, strong encounters 
in inducing bars.
(3)~We also take advantage  of the high resolution of the ACS ($\sim$~50~pc) to analyze a 
sample of 333 faint ($M_{\rm V}> -18$) dwarf galaxies in the Coma
core. Using visual inspection of unsharp-masked images, we find 
only 13 galaxies with bar and/or spiral structure. An additional eight
galaxies show evidence
for an inclined disk. The paucity of disk structures in Coma dwarfs suggests
that either disks are not common in these galaxies, or that any disks
present are too hot to develop instabilities.

\end{abstract}


\section{Introduction}\label{intro}

Mounting evidence  suggests that at  $z\sim$~1, major mergers
among massive galaxies are not very frequent (e.g., Bell et
al. 2006; Jogee et al. 2009; Robaina et al. 2009; Weinzirl
et al. 2009) and have not contributed significantly  to the
cosmic star formation rate (SFR) density (e.g., Bell et al. 2005;
Wolf et al. 2005; Jogee et al. 2009; Robaina et al. 2009)
or to the building of bulges in massive spirals (Weinzirl et al.
2009; Kormendy \& Fisher 2008; Laurikainen et al.~2007) since
$z<1$. Other processes, such as minor mergers  (Jogee et al.
2009; Weinzirl et al. 2009),  or internal secular processes
(Laurikainen et al.~2007;  Kormendy \& Fisher 2008) are
increasingly invoked.

The most efficient internal driver of evolution in
disk galaxies are stellar bars. They effectively redistribute angular momentum in the disk 
and dark matter (DM) halo, and drive large gas inflows to the central regions of galaxies. 
The resulting central dense gas concentrations ignite powerful starbursts (Schwarz 1981; 
Sakamoto 1999; Jogee 1999; Jogee et al.~2005; Sheth et al.~2005). In this way, bars are thought to
build disky central components known as pseudobulges
(Hohl 1975; Kormendy 1979; Combes \& Sanders 1981; Combes 1990; Kormendy 1993; Jogee 1999, Jogee et al.~2005).

We describe below some of the  recent progress made  in exploring 
the bar fraction  (defined as the fraction of disk galaxies hosting 
a large-scale bar)  in terms of methodology, dependence on  
Hubble type, redshift, and environment. 
For a long time, statistics on the  optical bar fraction in
the local universe came from visual classification of the galaxies in
the Third Reference Catalog of Bright Galaxies (RC3; deVaucouleurs et
al. 1991).  In the RC3, over all disk galaxies (S0--Im), the
visual optical bar fraction is $\sim$~30\% for strong bars (SB), 
and $\sim$~30\% for weak bars (SAB), giving $\sim$~60\% overall. 
The RC3 optical bar fraction suffers from two limitations: firstly, 
it denotes the optical bar fraction averaged  over a broad  range of 
Hubble types, and secondly, it is based on visual classification to 
identify and characterize bars, thus giving no quantitative
measurements of their properties (such as size, shape, strength).
In fact, the former limitation has persisted in many recent 
studies which  focus on the average bar fraction over
disks of all Hubble types (Eskridge et al. 2000; Knapen et al. 2000;
Hunt \& Malkan 1999; Mulchaey \& Regan 1997; Laine et al. 2002;
Laurikainen et al. 2004; Block et al. 2004; although see Odewahn
1996). 

Recent studies have made important headway on both fronts.
Firstly, quantitative methods such as ellipse 
fitting (e.g., Wozniak \etal 
 1995, Friedli \etal 1996; Regan \& Elmegreen 1997; Mulchaey \& Regan 1997; 
 Jogee \etal  1999, 2002, 2004; Knapen \etal 2000; Laine \etal 2002;    
 Sheth \etal 2003, 2008; Elmegreen \etal 2004; Men{\'e}ndez-Delmestre
 et al. 2007; Marinova \& Jogee et al.~2007, hereafter MJ07; Aguerri
 et al.~2009), and bulge-bar-disk decomposition (e.g.,
 Laurikainen et al.~2005; Reese et al.~2007, Weinzirl et al.~2009, Gadotti 2011)
are being used  to reduce the level of  subjectivity. 
Secondly, evidence is mounting that the bar fraction 
varies across galaxies of different Hubble types (Barazza, Jogee, \&
Marinova~2008, hereafter BJM08; Aguerri et al.~2009; Marinova et al.~2009, hereafter M09) and 
depends non-monotonically on the host galaxy properties, such as 
bulge-to-total ratio ($B/T$), luminosity, stellar mass, and color, at a range
of redshifts and environments (M09; Barazza et al.~2009; Weinzirl et
al. 2009; Laurikainen et al. 2009; Nair \& Abraham 2010a; Cameron et
al. 2010; Gadotti 2011).

Most of the results described above have focused
on  field galaxies at low redshifts, while 
much less is known about barred disks at higher redshifts 
(Abraham et al. 1999; Sheth et al. 2003; Jogee et al. 2004; Sheth et
al. 2008;  Cameron et al. 2010), 
and  in  dense environments.  
Dense environments such as clusters can be 
an important laboratory for studying the
co-evolution of bars and their host disks, 
as  there are many physical processes, which are unique to 
such environments and  impact disks and bars.
For example, we can study how the cumulative 
effect of frequent weak galaxy encounters (`harassment'; Moore et
al.~1996), 
the effect of the intra-cluster medium (ICM) on galaxies (Gunn \& Gott
1972; Larson et al.~1980; Quilis et al.~2000), and the
influence of the cluster potential as a whole (e.g.,  Byrd
\& Valtonen 1990; Gnedin 2003) affect the
properties of bars and disks. Theoretical studies 
exploring these processes give
conflicting predictions. For example, although the numerous tidal
encounters in a dense cluster core can induce bars in unbarred
disks in the case of a prograde encounter, they can also 
have little or no effect in terms of inducing a bar 
(or affecting the strength of an already existing bar) in the case of retrograde encounters (Gerin, 
Combes, \& Athanassoula 1990; Romano-D{\'{\i}}az et al.~2008; Aguerri \&  Gonz{\'a}lez-Garc{\'{\i}}a 2009).
 Determining the fraction and properties of barred galaxies in
such extreme environments and comparing them to those of galaxies in
environments of varying densities, can give clues to the outstanding problems
in understanding the formation and growth of bar instabilities in disks, and
therefore to understanding disk evolution. In addition,
identifying barred disks in cluster environments can provide a lower
limit to the fraction of disk systems in clusters (e.g., M09; M{\'e}ndez-Abreu et al. 2010).

Quantitative results
addressing this issue are only starting to emerge. Several recent
studies of non-dwarf galaxies at a range of redshifts from $0.01 < z < 0.4$ 
(Barazza et al. 2009; Aguerri et al.~2009; M09), find that the optical bar fraction shows at most a modest variation 
($\pm10\%$) between field and intermediate density environments (e.g., moderately rich clusters). 
But what happens at really high
densities (i.e., dense cluster cores)? Some studies (Thompson 1981;
Andersen 1996; Barazza et al. 2009) have suggested that, within a galaxy cluster, the
bar fraction may be higher in the dense core regions than the outskirts. 
However, this has remained an open question for cluster cores due to issues
such as limited number statistics, projection effects, poor quality or inhomogeneous data, 
 and  uncertainties in cluster membership.  We make progress in two
 ways
with this study. Firstly, we are able to establish cluster membership 
for all galaxies in our bright S0 sample (see $\S~\ref{disks}$) using
spectroscopic redshifts. Although we are still limited by
line-of-sight projection effects from within the Coma cluster itself, 
the number density of galaxies in the core of
Coma is a factor of 10-100 times higher than at larger cluster radii, so 
we expect the impact of such line-of-sight contamination to be
low. Secondly, the high-resolution $HST$ ACS data is vital in identifying bars in S0s, especially in the
types of cases we discuss in Section~\ref{barred}, namely, where a relatively 
large bulge, in combination with a relatively short bar and projection 
effects combine to make bar detection difficult in low-resolution 
ground-based images. 

In this paper, we provide baseline results for the 
densest environment in the local Universe: the 
central regions of Coma. 
We conduct two explorations using two different
samples. In the first part of this paper ($\S$~\ref{disks}), we focus on the sample of
bright ($M_{\rm V} \lesssim -18$) non-dwarf S0 galaxies, which comprise 94\% of 
our sample of bright disk galaxies in the Coma core. We characterize their
disk and bar properties and present results on the optical bar fraction 
for bright S0 galaxies from ellipse fitting and visual classification.  
We compare with the results of other
studies of S0 galaxies in Coma, and less dense clusters (Abell 901/902 and Virgo).

In the second part of this paper ($\S$~\ref{dwarfs}), we take advantage of the exquisite
50~pc resolution of the ACS images to search for bars and other disk
features (spiral arms, inclined disks) in the faint, dwarf ($M_{\rm V} > -18$) galaxies of the Coma cluster core. 
The prevalence or paucity of bar
structures in early-type dwarf galaxies can not only provide clues on the
evolutionary history of these dwarf systems, but also has implications
for the conditions necessary for bar formation and growth in
galaxies. Are bar/spiral arm instabilities commonplace for dwarf galaxies? 
Some early-type dwarfs in clusters are believed to have originated from late-type spirals or
dwarf irregulars (e.g., Lin \& Faber 1983; Kormendy 1985) that have been
stripped of their gas by external processes. This is
 especially true in a cluster setting where environmental processes like
ram-pressure stripping and harassment are commonplace.  
If the progenitor galaxy hosted a stellar bar or spiral arms, these stellar features
can persist even after the galaxy's gas has been processed by the cluster environment.
In fact, a number of early-type dwarf galaxies in Virgo have been
observed to host stellar disk features such as a
lens, a bar, or spiral arms (e.g., Sandage \& Binggeli 1984; Binggeli \&
Cameron 1991; Jerjen et
al. 2000; Barazza et al. 2002; Lisker et al.~2006; Lisker et al.~2007;
Lisker \& Fuchs 2009) thus supporting the above scenario. 
Graham et al.~(2003) discovered the first two early-type dwarf galaxies with
spiral-like structure in the Coma cluster using unsharp masking, 
concluding that such galaxies may provide the `missing link' in the evolution 
of faint spiral galaxies into dwarf spheroidals due to
cluster processes. 

The layout of this paper is as follows: 
In $\S$~\ref{data} we describe our
dataset and sample selection. Section~\ref{disks} deals with our 
bright ($M_{\rm V} \lesssim -18$) Coma core sample. In this section we outline our methods
for identifying bright S0 galaxies ($\S$~\ref{nondwarfget}), our
methods for identifying bars in these galaxies ($\S$~\ref{diskmethod} and \ref{visual}),
our results for S0 galaxies in the Coma core
($\S$~\ref{fbarS0coma}--$\S$~\ref{barredpropdisc}), and the implications
of these results for bar and disk formation and evolution in dense
environments ($\S$~\ref{diskdisc}). 
In $\S$~\ref{dwarfs}, we present our investigation of the faint, dwarf galaxies 
in our Coma core sample. We describe our methods for finding disk
features (bar, spiral arms, inclined disk) in these galaxies
($\S$~\ref{dwarfmethod}), as well as our results and discussion for
dwarf galaxies in
$\S$~\ref{dwarfdisc}. 
We summarize all of our results in $\S$~\ref{summary}. 

\section{Data and Selection of a Cluster Sample}\label{data}
Our data come from the \textit{Hubble Space Telescope} ($HST$) Advanced
Camera for Surveys (ACS) Treasury survey of the Coma 
cluster at $z\sim$~0.023 (Carter et al. 2008). Originally designed to
cover a large area of the core and infall region of Coma, the survey
 remains only $\sim$~28\% complete because of the failure of ACS in
 2007. Nevertheless, the available data span 274 arcmin$^{2}$, where
 approximately 75\% of the data are within 0.5~Mpc of the
 cluster center. The data cover approximately 70\%
of the core region of Coma (assuming $R_{\rm core} \sim$~0.2~Mpc) and
are therefore representative of the core and 
the immediate surroundings, namely the region where the galaxy number density
is $\sim$~10,000 galaxies Mpc$^{-3}$ before it drops 
sharply as a function of distance from the cluster center (The \& White 1986). 
This dataset contains thousands of
 sources down to a limiting magnitude of $I =$~26.8 mag in F814W (AB
 mag).
The ACS point-spread function (PSF) is $\sim0.1\arcsec$, which
 corresponds to   $\sim$~50~pc at the distance of the Coma cluster,  assuming  
a distance of $D = 100$~Mpc.\footnote{We assume in this
paper a flat cosmology with $\Omega_M = 1 - \Omega_{\Lambda} = 0.3$
and $H_{\rm 0}$ =70~km~s$^{-1}$~Mpc$^{-1}$.}. 
$SExtractor$ source catalogs are available as a part of the
 Coma survey data releases. The second data release (DR2)
 is described in detail in Hammer et al.~(2010). Throughout
the paper, $M_{\rm I (814)}$ as well as SDSS $g$  and $r$ magnitudes are given in the AB system, 
while $B$ and $V$ magnitudes are in the Vega system.

We first use the eyeball catalog of Trentham et al. (in preparation) to
select cluster members. In this catalog, galaxies
are visually assigned a cluster membership class from 0 to 4. Galaxies
with membership class 0 are spectroscopically confirmed members, while
galaxies with class 4 are visually deemed to be likely background
objects. The intermediate membership classes from 1 (very probable cluster
member) to 3 (plausible cluster member), are assigned based on a visual estimation taking into account both surface
brightness and morphology (Trentham et al. in prep). 
We select objects with apparent magnitude $m_{\rm I (814)}\le$~24~(AB mag) and membership class 0 to 3, 
resulting in a sample of 469 cluster galaxies. For these galaxies, 41\%
are spectroscopically confirmed members (class 0), while 7\%, 27\%,
and 25\% have membership classes 1 to 3, respectively. 

We derive $B$ and $V$ magnitudes (in Vega mag) for the Coma galaxies using SDSS 
$g$ and $r$ (in AB mag). For the bright sample, we use the $B$, $V$, $g$, and $r$ magnitudes
 (instead of the ACS F814W) for ease of comparison to other studies.  In addition, Hammer et al. (2010)
find that for bright galaxies ($M_{\rm I (814)} \le 17$~AB mag) in the Coma survey, it is more 
reliable to use the SDSS rather than ACS magnitudes, as the latter may be 
unreliable for some galaxies with large, diffuse stellar halos.  
We use the following transformations from Jester et al.~(2005) to 
convert the SDSS $g$ and $r$ (AB) to $B$ and $V$ (Vega)\footnote{The transformation
equation tables can be found at http://www.sdss.org/dr7/algorithms/sdssUBVRITransform.html}:
\begin{equation}
B      =    g + 0.39\times(g-r) + 0.21    
\end{equation}
\begin{equation}
V      =    g - 0.59\times(g-r) - 0.01. 
\end{equation}
We calculate absolute magnitudes assuming a distance modulus of 35.0 (Carter et al.~2008). 

In this paper, we explore the optical bar fraction in two regimes: 
bright, non-dwarf S0 disks ($\S$~\ref{disks}) and faint (dwarf) galaxies ($\S$~\ref{dwarfs}). 
To separate these two regimes, we apply a magnitude cut of 
$M_{\rm I (814)} \le -18.5$, roughly 
equivalent to $M_{\rm V} \lesssim -18$ or $M_{\rm B} \lesssim -17$ for our sample.
We choose a cut at $M_{\rm V} \sim -18$ because it tends to separate well the regimes 
where normal and dwarf galaxies dominate on the luminosity functions of clusters (Binggeli et al. 1988; Trentham 1988; 
Trentham \& Hodgkin 2002; Mobasher et al. 2003).
Luminosity cuts at magnitudes $ M_{\rm V} \sim -18$ are often used to separate
dwarf and non-dwarf galaxies in the literature (Matkovi{\'c} \& Guzm{\'a}n 2005; Aguerri et al. 2005; 
van Zee 2001; Barazza et al. 2006).
This cut gives 52 galaxies brighter than, and 417 
galaxies fainter than 
$M_{\rm V} = -18$. 
From the 52 bright galaxies, we discard four galaxies (two are a close/merging pair and two galaxies are partially off
the edge of a tile) bringing our initial bright sample to 48 galaxies. 
We discuss the methods for selecting S0 galaxies from this bright 
sample in $\S$~\ref{nondwarfget}. 

In Fig.~\ref{totpropbf}a we show the absolute magnitude $M_{\rm I (814)}$ distribution
for the non-dwarf (bright) and dwarf Coma core samples. Fig.~\ref{totpropbf}b shows the 
peak surface brightness $\mu_{\rm 0}$ (from the $SExtractor$
source catalogs in DR2) vs. absolute magnitude $M_{\rm I (814)}$ distribution
of the bright, non-dwarf and dwarf galaxies.

\section{Bars in bright S0 galaxies in the central region of the Coma cluster}\label{disks}
\subsection{Identifying Bright S0 Galaxies}\label{nondwarfget}
Due to the fact that bars are inherently a disk phenomenon, the bar fraction
is traditionally defined as 
\begin{equation}
f_{\rm bar} = \frac{N_{\rm bar}}{N_{\rm bar} + N_{\rm unbar}},
\end{equation}
 where $N_{\rm bar}$ and $N_{\rm unbar}$ represent 
the number of barred and unbarred disk galaxies, respectively. 
Therefore, from the bright sample of 48 galaxies, we need a sample
of disk galaxies (e.g., S0--Im) for analysis.   As discussed in $\S$~\ref{intro}, recent work has shown that a bar
fraction averaged over a wide range in Hubble types gives only limited
information. The bar fraction is a strong function of galaxy
properties, such as $B/T$, luminosity, stellar mass, and color.   
Because our sample is too small to split into fine bins by
morphological type, and because most (94\%) of bright 
disk galaxies in our Coma core sample are S0s (see below), \textit{our analysis of bright galaxies in this paper focuses on S0s only}. 
The goal of our study is to provide the bar fraction for the 
densest low redshift ($z\sim$~0.02) environment and 
to serve as a comparison point for 
studies of barred S0 galaxies in field and intermediate-density environments 
at different redshifts.  

Starting with the bright sample of 48 galaxies, we use visual classification 
to separate the galaxies into ellipticals, S0s, visually ambiguous E/S0, and 
spirals. We note that our visually-identified class of `S0' 
galaxies includes all Hubble type S0 sub-types from S0$^{-}$ to S0/a (numerical T-types -3 to 0). 
It is fairly easy to separate S0s and ellipticals visually when 
the S0s host bars. However,  
unbarred S0s are harder to separate from ellipticals since an unbarred S0 hosts
a disk, which is effectively featureless and devoid of tell-tale disk signatures, such as 
a bar or spiral arms. We find a 
group of 10 bright galaxies that are \textit{visually} ambiguous E/S0s. 
For each of these 10 galaxies, we perform multiple component
structural decomposition with the  GALFIT code (Peng et al.~2002)
by fitting  the two-dimensional (2D) light distribution  taking into
account the PSF, following the procedure described in Weinzirl et
al.~(2009).
In brief, we fitted each galaxy with three models: a single-component 
S\'ersic model, a  two-component (`bulge'+disk)  model,  and a 
three component  (`bulge'+disk+bar) model.  
The disk is represented by an exponential  (S\'ersic $n =1$) model, 
while the `bulge' is a S\'ersic model with a free-floating S\'ersic
index $n$.  The bar is typically associated with a S\'ersic  model
of low $n$. If needed, a point source 
component was added to represent the point sources, which are common
in the center of these galaxies.
  For each model,
GALFIT finds the optimum solution using the Levenberg-Marquardt algorithm.
The goodness of fit is determined iteratively by calculating $\chi^{2}$.
GALFIT continues to adjust the model parameters until the gradient $\delta \chi^2 /
\chi^2$
is very small (e.g., 10$^{-4}$) for 10 continuous iterations.
Out of the three models (S\'ersic, `bulge'+disk, `bulge'+disk+bar), the 
best one was selected by considering a number of factors, including the
$\chi^{2}$ values, the strength and spatial distribution of the
residuals, and the physical viability of output parameters   (e.g., 
effective radii of the S\'ersic and bulge component; scale length,
axial ratio, and position angle of the disk). 
 These factors were used to decide whether  the galaxy
is likely an elliptical or an S0. Examples of the data, model, and
residuals for three representative galaxies classified as E, S0/a, and
E/S0 after decomposition are
shown in Figure~\ref{bdmontage}.

Out of the 10 \textit{visually} ambiguous E/S0 galaxies, we 
find from multi-component decompositions that one is an elliptical,
eight are S0 or S0/a, and one is still ambiguous E/S0.
 The ambiguous galaxy may be 
a disk galaxy with complex structure or an elliptical with an inner 
debris disk. In $\S$~\ref{fbarS0coma} we use this galaxy to 
estimate the uncertainty in the optical bar fraction
by calculating $f_{\rm bar}$ (see Equation 1) for the two cases where this 
galaxy is either included or excluded in the number of unbarred 
disks ($N_{\rm unbar}$). 

The absolute magnitude $M_{\rm I (814)}$ distribution 
of the bright Coma core sample is 
shown in Fig.~\ref{totprop}a.  
The final morphological breakdown of our bright sample (13 ellipticals, 
1 ambiguous E/S0, 32 S0s, and 2 spirals) is shown in Fig~\ref{totprop}b. It 
is clear that S0s dominate among the bright disk galaxies in our 
Coma core sample, which is expected for the central 
regions of a dense cluster.  We find a ratio of E : S0 : Sp of 
28\% : 68\% : 4\%. This is at the extreme end of the morphology-density
relation found in dense environments by Dressler (1980). 
Fig.~\ref{totprop}c shows the distribution of
stellar mass for the S0 disk sample 
as well as all bright ($M_{\rm V} \lesssim -18$) 
galaxies. Stellar masses are calculated using the relations 
from Borch et al.~(2006)\footnote{The Kroupa IMF offset term is
  reported as -0.15 in Bell et al.~(2003). However this value was
  calculated assuming unrealistic conditions (Bell, E., private communication). The correct value of
  -0.1 was recalculated and reported in Borch et al.~(2006).} assuming a Kroupa et al.~(1993) initial mass function:
\begin{equation}
\frac{M}{M_{\odot}} = v_{\rm lum} \times 10^{-0.628 + 1.305(B-V) - 0.1},
\end{equation}
where
\begin{equation}
v_{\rm lum} = 10^{-0.4(V - 4.82)}.
\end{equation}
Galaxies in our S0 disk sample have stellar 
masses between 10$^{9.5}$ and 10$^{11} \msun$.
Fig.~\ref{totprop}d shows a $g-r$ color vs. 
$M_{\rm r}$ magnitude diagram. Almost all
ellipticals and most disk
galaxies fall on the red sequence. We overplot the
relation from Blanton et al.~(2005a) for the break between the red sequence and blue cloud using the
equation (modified with an offset of -0.77 for $h=0.7$)
\begin{equation}
(g-r)=0.65-0.03(M_{\rm r}-0.77+20).
\end{equation}
We also plot a subsample of the dwarf galaxies ($M_{\rm V} > -18$)
for which SDSS data are
available ($\sim$~30\%). 

We show examples from our final bright S0 
sample of 32 galaxies in Figure~\ref{s0montage}. We note that 
all 32 S0s in the bright sample are spectroscopically confirmed cluster members.

\subsection{Identification of bars in S0s via ellipse fits}\label{diskmethod}

Ellipse fitting is our primary method of detecting bars in the bright S0 sample (e.g., Wozniak \etal 
 1995, Friedli \etal 1996; Regan \etal 1997; Mulchaey \& Regan 1997; 
 Jogee \etal  1999, 2002, 2004; Knapen \etal 2000; Laine \etal 2002;    
 Sheth \etal 2003, 2008; Elmegreen \etal 2004; Men{\'e}ndez-Delmestre et al. 2007; MJ07; Aguerri et al.~2009). 
To detect bars through ellipse fitting we use the
standard IRAF task $ELLIPSE$ in conjunction with an adaptive wrapper (Jogee et
al. 2004), which runs $ELLIPSE$ iteratively on each galaxy until the
best fit is found or up to a maximum number of times specified by the
user. Ellipses are fit to the galaxy isophotes out to a maximum
distance ($a_{\rm disk}$) where the brightness of the isophotes reaches
the noise level. We note that the value of $a_{\rm disk}$ depends
on the depth of the image, as $a_{\rm disk}$ will reach 
larger values for deeper images. However for the purpose of bar 
detection, it is only necessary for the radial profile to 
extend beyond the bar into the more circular region of the disk. 
We typically set the  maximum allowed iterations to 300,
however for most galaxies a good fit is achieved in only a few
iterations. A good fit is one where an ellipse can be fitted at
every isophote out to $a_{\rm disk}$. Residuals characterizing how
well each isophote is fitted by its corresponding ellipse are given by
the harmonic amplitudes A3, B3, A4, and B4 (e.g., Carter 1978; Jedrzejewski 1987; Carter 1987). 
For our barred galaxies, we find typical amplitudes of 5--10\%. For a detailed
discussion on the advantages and drawbacks of using ellipse fitting to
characterize bars we refer the reader to MJ07. 

Once the galaxies are fitted, we use an interactive visualization tool
 to display the overlays of the fit on the galaxy
image, as well as the radial profiles of surface brightness,
ellipticity ($e$), and position angle (PA). Using the radial profiles 
of ellipticity ($e$) and PA, we classify the galaxies as `highly inclined', `barred', or
`unbarred'. We discuss these classes in more detail below. 

\subsubsection{Detecting and removing highly inclined galaxies}\label{inclined}
In studies of bars, it is conventional to exclude highly inclined galaxies as 
the large inclination precludes accurate structural classification, 
making it particularly difficult to identify systems as barred or unbarred. 
We use two ways to identify highly inclined galaxies in the bright S0 sample. 
The first is via the ellipse fit criteria, where the observed
outermost disk
isophote (at $a_{\rm disk}$)  has $e > 0.5$, corresponding to $i >
60^{\circ}$. We find eight galaxies that fit this criterion. 
This method works well for spirals of intermediate to late Hubble 
types, but does not capture all highly inclined disks for S0s because 
for some edge-on or highly inclined S0s, a rounder, thickened
outer stellar component can sometimes dilute the outermost isophote so that the
outermost ellipticity is below 0.5 although the galaxy is highly inclined.

Therefore, our second method is to visually identify highly-inclined S0s. 
We visually identify these systems using the criteria
that a thinner, high-surface brightness, highly-inclined 
disk appears embedded in a thick, diffuse stellar component, which
 could be a mix of thick disk, bulge, and bar stars. 
A typical example of one of these galaxies is shown in Fig.~\ref{visincexample}.
We encounter four such 
cases, where the 
galaxy appears visually to be close to edge-on, and is classified as highly inclined. 
 We distinguish 
these cases from a face-on galaxy with a bar because (1)~the 
thin, highly-inclined disk and the thicker stellar component are always oriented along the 
same position angle, (2)~the thick outer stellar component in these highly-inclined S0 galaxies 
appears much fainter and more
diffuse than a face-on disk, and (3)~in three out of these four galaxies, a box or X-shaped 
bulge is present, suggesting that the galaxy is seen edge-on (e.g., Athanassoula 2005). 
We therefore exclude from further analysis the 12 highly-inclined systems that we find in the sample. 

\subsubsection{Detecting barred galaxies}\label{barred}
Traditionally, when ellipse fits are used to identify bars, 
a galaxy is classified as barred if the radial profiles
of the ellipticity and PA fulfill the following
requirements:~(1)~the $e$ rises to a \textit{global} maximum,
$e_{\rm bar} > 0.25$, while the PA remains relatively constant (within $\pm 10^{\circ}$), 
and (2)~the $e$ drops by at least 0.1 and the PA 
changes by more than 10$^{\circ}$ at the transition between the bar and disk region. An 
example of a barred S0 galaxy in our sample that meets the traditional 
criteria is shown in  Fig.~\ref{ellipseex1}.

These criteria will identify primary stellar bars in the vast majority
of spirals, particularly those of intermediate to late Hubble
types (Sb--Sm).  However, they can marginally fail in some galaxies
due to a 
rare set of circumstances, which we describe in detail
below. These circumstances are particularly likely 
to occur in S0s with large bulge-to-disk ratios.

In some S0s, a combination of structural parameters and viewing angle 
causes the observed (i.e., not deprojected)
maximum bar ellipticity ($e_{\rm bar}$; typically measured from 
the isophote crossing the end of the bar)  to  become a {\it local}  maximum of the ellipticity radial 
profile rather
than the {\it global} maximum. In these rare cases, 
the ellipticity ($e_{\rm disk}$) of the  outer disk 
becomes the {\it global} maximum in the radial profile of ellipticity.
This can happen in the case of a barred galaxy, where
all or  most of the following conditions are satisfied:
(i)~The galaxy has a moderate to large inclination (e.g.,
$i > 50^{\circ}$).
  This causes the outer circular disk of the galaxy to appear
   elongated along the line of nodes (LON) in the projected image
   of the galaxy on the sky, leading to a higher measured
   $e_{\rm disk}$.
(ii)~A large fraction of the length of the bar lies
within a fairly axisymmetric bulge, which is much 
more luminous than the bar. In this case, the bulge light dilutes the
    ellipticity of the bar by `circularizing' the isophote
crossing the bar end, thus causing $e_{\rm bar}$,
   measured from this isophote to be significantly
   lower than the true ellipticity of the bar. 
(iii)~The bar major axis has a large offset ($\Delta\theta$) with
    respect to the LON,  such that projection effects make the disk appear more 
elongated, while the bar appears more round. The most extreme example occurs
when the bar is perpendicular to the LON (i.e., $\Delta\theta = 90^{\circ}$).
 Such situations can potentially cause
    the observed ellipticity of the disk to exceed that of the bar.
Thus, a combination of factors (i) to (iii) can cause the
measured $e_{\rm bar}$ to fall below $e_{\rm disk}$
so that $e_{\rm bar}$ is a {\it local} maximum in the $e$ radial 
profiles.  In this case,
the bar can still be identified through ellipse-fits if  the
traditional criterion that  the measured maximum bar ellipticity
$e_{\rm bar}$ must be a {\it global} maximum is relaxed,
and a {\it local} maximum be deemed acceptable. 

In the case of barred S0s,
the conditions (i) to (iii)  can be satisfied in a larger fraction
of galaxies  than for a sample of  barred intermediate-to-late
Hubble type  (Sb--Sm)  spirals due to the following reasons.
Many barred cluster S0s host bulges that are bright,  have
large  bulge-to-disk  light ratios, and
encompass a large fraction of the length of the bar. Indeed, among
our sample of 20 moderately-inclined cluster S0s,  we find three such cases and an
example is shown in  Fig.~\ref{ellipseex2}. For this reason, 
we quote two bar fractions derived through ellipse fits:
 the first bar fraction ($f_{\rm bar, ES}$), where we use the strict criteria (1) and (2) above, 
and the second bar fraction ($f_{\rm bar, ER}$) where for galaxies satisfying (i) to (iii),
we relax the criterion that the maximum bar ellipticity
must be a {\it global} maximum (however we still require it to rise 
above $e = 0.25$). We note that all 
bars identified with the strict ellipse-fitting criteria (`ES') are 
also picked up under the relaxed ellipse-fitting criteria (`ER').  
We further note that, for the galaxies where the bar is detected only 
through the relaxed criteria,
 if the radial profiles of the ellipticity and PA are deprojected ($\S$~\ref{barredpropdisc}), 
the bar ellipticity $e_{\rm bar}$ does become the global maximum. 
However, because many large studies of bars do not deproject the 
radial profiles (e.g., M09), we use the observed radial profiles 
to detect the bars as described above for ease of comparison.

\subsection{Identification of bars in S0s via visual classification}\label{visual}

In addition to ellipse fitting, we also present the optical
bar fraction for bright S0s in Coma from visual classification performed
by I.M., S.J., and P.E. This facilitates comparison to other work where
bars are identified visually ($\S$~\ref{s0comp}).  

A galaxy is classified as `barred' through visual classification
if it has a significant elongated feature extending from the center of 
the disk with an axial ratio (estimated from the image with $DS9$) $< 0.7$ 
and a PA that differs from the PA of the outer disk 
by at least 10$^{\circ}$. 
A galaxy is classified as `unbarred' if there is no 
elongated structure present that fits the above criteria. 


All of the bright barred S0s we identify in the Coma sample through 
ellipse fitting and visual classification are listed in 
Table~\ref{barinfo}. The methods through which the bar is detected
are shown in column (5).

\subsection{Optical S0 bar fraction in the central region of the Coma cluster}\label{fbarS0coma}
The optical bar fractions for our sample of bright S0 galaxies in the 
Coma cluster core are presented in Table~\ref{bar_stats}. 

Using the strict ellipse fitting criteria ($\S$~\ref{barred}), 
we find  that the optical bar fraction for the bright S0s is $f_{\rm bar, ES}=$~50~$\pm$~11\% (10/20). Using
the relaxed ellipse fitting criteria, we find $f_{\rm bar, ER}=$~65~$\pm$~11\% (13/20). Visual 
classification gives an optical bar fraction of 60~$\pm$~11\% (12/20).
All errors are binomial errors. 
The barred S0 galaxies identified through ellipse fitting and visual
classification  are shown in 
Fig.~\ref{barstamps}.

To correctly derive the bar fraction for S0s in clusters,
we need to accurately estimate the number of
unbarred S0s ($N_{\rm unbar}$ in Eq.~1).  
As S0s are devoid
of typical disk features such as spiral arms,  star-forming
rings, etc., it is particularly challenging to {\it visually}
identify all unbarred S0s and separate them from ellipticals.
Therefore, in $\S$~\ref{nondwarfget}, we identified S0s through both
visual classification and two-dimensional structural
decomposition of the images into single-component
S\'ersic models, bulge+disk models, and  bulge+disk+bar
models. We found  13 Es, 32 S0s, and one  E/S0 case, which
still remains ambiguous even after
decomposition.  This ambiguous E/S0 case is not included in
the optical bar fraction in Table~\ref{bar_stats}. 
If it is included as an
unbarred S0 in our analysis, the optical bar fractions fall
to:  $f_{\rm bar, ES}=$~48$\pm$~11\%, $f_{\rm bar, ER}=$~62$\pm$~11\%, 
and 57$\pm$~11\% from visual classification. 
 We therefore  estimate that the uncertainties associated
with determining the number of unbarred S0s can lead us to
overestimate the optical bar fraction by only a small factor 
of $\sim$~1.05. 

We note that in the field, the bar fraction is lower
in the optical than in the NIR by factor of 1.3 (Eskridge et al.~2000; M07)
for intermediate (Sab--Sc)  Hubble types due to obscuration by gas, dust,
and star formation. However
in S0s, where there is little gas and dust on large scales, we don't 
expect the
 difference between the optical and NIR bar fractions to be significant.

\subsection{S0 bar fraction across different environments}\label{s0comp}

Due to the fact that different bar detection methods can yield 
different bar fraction results, it is important to compare studies
using the same methods for consistency. 
In addition, as discussed in $\S$~\ref{intro}, the bar fraction 
depends on host galaxy properties such as Hubble type or $B/T$
(Odewahn 1996; BJM08; Aguerri et al.~2009; M09; Weinzirl et al.~2009; Laurikainen et al.~2009),
luminosity (Barazza et al.~2009; M09), stellar mass, and color (Nair \&
Abraham 2010a; Cameron et al.~2010). 
Therefore we use comparison samples
that are matched as well as possible to our Coma sample 
in Hubble type (S0s), luminosity ($M_{\rm V} \lesssim -18$), color,
and method of bar detection.
We compare our results for S0s to those of other studies in 
Coma and lower-density clusters (Abell 901/902 and Virgo).

First, we compare to 
another study in the very dense environment of the
central regions of the Coma cluster (galaxy number
density $n \sim$~10,000 galaxies/Mpc$^{3}$) by T81.
T81 uses visual classification on ground-based Kitt Peak National Observatory
(KPNO) plates to detect bars in S0s 
brighter than $M_{\rm V} = -17.5$ (very similar to our magnitude
cutoff of $M_{\rm V} \sim -18$).  
Therefore we compare his result to our optical 
bar fraction from visual classification ($\S$~\ref{visual}).
 Table~\ref{envt_comp} shows that our optical bar fraction from 
visual classification for S0s
(60$\pm$11\%) in the
Coma core is higher than the result (42$\pm$7\%) that
T81 obtained after correcting raw galaxy counts for projection
effects\footnote{This correction is used by T81 to account for the
  effects of foreground and background objects contaminating the
  cluster field in the absence of spectroscopic data. Since all bright S0s in our Coma core sample are
  spectroscopically-confirmed
cluster members, we compare to the  corrected bar fraction from T81.}. 
A clue to the reason for this difference comes from our finding 
that in many S0s in our sample, the bar ellipticity and its overall 
signature are diluted because the bulge is bright compared to the bar 
and it is large enough to encompass a large fraction of the bar length. 
In such cases, as discussed in section $\S$~\ref{barred}, the bar is harder to 
detect via any method, be it visual classification or ellipse fits,
unless the image is of high quality and the classifier has 
significant expertise. In the case of T81, the visual classification
 was performed on ground-based optical plates, which are of lower quality 
 than CCD images, making it even more difficult to detect such diluted or/and 
 short bars. It is not possible to directly compare our case-by-case 
results with T81, as he does not publish the list of galaxies he classifies
as barred and does not provide the lengths and ellipticities of the bars.
However, we perform two indirect tests to gauge the impact of 
 missing diluted, weak, and/or short bars.
In our sample, seven of the 13 barred S0s have 
an observed peak bar ellipticity $e_{\rm bar} < 0.4$ (Fig.~\ref{abarebar}). 
 If all of these galaxies were classified as unbarred, 
the bar fraction would drop to 30\%.  
Alternatively, if the shorter ($a_{\rm bar} < 2$~kpc) bars 
are excluded, the bar fraction would drop to 45\%. These tests suggest that it 
is likely that the lower optical bar fraction of T81 is due to his missing
some of these diluted or/and short bars.

For a comparison to intermediate-density cluster environments, 
we use the study of M09 for the Abell 901/902 cluster system
($z\sim$~0.165; $n \sim$~1000
galaxies/Mpc$^{3}$, see Table~5 in Heiderman et al.~2009).
To match our sample, we
pick S0 galaxies from the M09 study with $M_{\rm V} \le -18$, using classifications
performed by the members of the STAGES collaboration for the Abell 901/902 cluster system (Gray et
al. 2009; see Wolf et al. 2009 for more details). In M09,  
inclined galaxies were picked as those with 
outer disk ellipticity $e_{\rm disk} > 0.5$, as traditionally done in bar studies using 
ellipse fitting. However, since we are only focusing on S0 galaxies (which
sometimes have large bulges/diffuse, thick stellar components that can dilute the $e_{\rm disk}$
below 0.5 even for edge-on S0s, as discussed in $\S$~\ref{inclined}) we apply the same additional visual criteria
outlined in $\S$~\ref{inclined} to the M09 S0 sample to detect and 
remove highly inclined S0s. 

Figure~\ref{m09sampprops} shows the host galaxy properties of the
 Abell 901/902 and Coma S0  samples. The two samples are well-matched in 
the {\it mean}  luminosity, $g-r$ color,  and stellar mass. However, the 
results  of a Kolmogorov-Smirnov (KS) test show differences in the 
overall distributions of $g-r$ color and stellar mass 
(KS ($p = 0.002$, $D=0.4$) and (0.005, 0.3), respectively),  and are inconclusive 
for the distributions of $M_{\rm V}$ (KS ($p = 0.13$, $D=0.2$)). This is
likely due to the fact that the Abell 901/902
sample has a tail of galaxies with masses both lower and higher than 
the Coma core sample, translating, respectively, into a 
tail of bluer colors and brighter absolute magnitudes.
Oddly, there appears to be a significantly larger offset between
the Abell 901/902 cluster and the Coma core if one uses the 
$B-V$ color (Figure~\ref{m09sampprops}b) rather than the $g-r$ 
color  (Figure~\ref{m09sampprops}c):  Abell 901/902 S0s appear $\sim$~0.2
mag bluer in $B-V$ color compared to the Coma core, and the KS test 
suggests a large difference  (KS ($p = 6*10^{-16}$, $D= 0.8$)).
However, we believe that this $B-V$ color offset  is not  real
and is  caused by the fact that the color transformations derived 
by Jester et al. (2005)  for stars may not be  adequate for S0s.

In M09 bars were
detected on optical ($HST$ ACS F606W) images via the strict ellipse-fitting criteria 
only. We therefore also derive the bar fraction through visual classification 
and using the `relaxed' ellipse-fitting criteria for the S0s in 
the M09 sub-sample in order to derive the corresponding bar fractions for comparison
to Coma S0s. As shown in 
Table~\ref{envt_comp}, we find no statistically significant difference
in the S0 bar fraction in Abell 901/902 and Coma clusters when detecting bars
through visual classification, strict ellipse fit criteria, or relaxed
ellipse fit criteria. 
For instance, from visual classification (Table~\ref{envt_comp}, bottom section), 
the Coma bar fraction (60$\pm$11\%) is slightly 
higher than the A901/902 bar fraction (55$\pm$5\%), but the 
difference is not statistically significant as the values
are consistent within the error bars.
Similarly, via the relaxed ellipse fitting criteria 
(Table~\ref{envt_comp}, middle section), the Coma bar fraction is 65$\pm$11\%, while 
the A901/902 bar fraction is 48$\pm$5\%, (barely) consistent within the
error bars. 
A comparison of the bar and disk properties
(such as $e_{\rm bar}$ and $a_{\rm bar}/R_{25}$) in Coma and the 
Abell 901/902 cluster system is discussed in
$\S$~\ref{barredpropdisc}. 

Next, we compare to results in Virgo from Erwin et al. (in
preparation; E11). Virgo is the 
most nearby cluster ($D\sim$~20~Mpc, $z\sim$~0.005) and is representative of a low-density
cluster environment ($n \sim$~300 galaxies/Mpc$^{3}$ in the core region). We note
however, that different environmental tracers paint different pictures
in Virgo. While the number density ($n \sim$~300
galaxies/Mpc$^{3}$) is lower than that of Abell 901/902 ($n \sim$~1000
galaxies/Mpc$^{3}$) or the Coma core ($n\sim$~10,000
galaxies/Mpc$^{3}$), the velocity dispersions in Virgo can be as high 
as 750~km/s (Binggeli et al.~1987), comparable to those seen in Abell
901/902 and much higher than in groups ($\sim$~100~km/s; Tago et
al.~2008). These properties are relevant for the discussion of our
results in $\S$~\ref{diskdisc}.
Our Virgo comparison sample consists of S0 galaxies brighter than
$M_{\rm V} = -18$ from E11. 
Fig.~\ref{pe11sampprops}c shows that the two  samples agree well in 
 $g-r$ color (KS ($p = 0.3, D = 0.4$)), but Fig.~\ref{pe11sampprops}b 
shows that the $B-V$ 
colors of the Virgo S0s are  significantly bluer (by $\sim$~0.15~mag, on 
average, and with KS values of ($p = 8 * 10^{-11}$, $D=0.9$)). We believe 
that this $B-V$ color offset  between  Coma and Virgo  is not  
real since the measured $g-r$ colors for  the two  samples agree well. 
We again think this offset is likely caused by the possibility
that the color transformations derived by Jester et al. (2005)  
for stars may not be  adequate for S0s.
We compare optical bar fractions derived
with all three methods: strict ellipse fitting criteria (ES), relaxed
ellipse fitting criteria (ER), and visual classification performed by P.E., I.M., and S.J. 
according to the criteria outlined in $\S$~\ref{visual}. Again, we do not
find a statistically significant difference in the optical bar fraction (within the
errors) for S0s in the Coma core and those in Virgo using any of the three
bar-detection methods above (see Table~\ref{envt_comp})\footnote{We
  note that Giordano et al.~2010 quote a much lower bar fraction
  ($\sim$~30\%) using visual classification for Virgo S0
  galaxies. This lower value is likely due to the fact that Giordano
  et al.~2010
  include much fainter galaxies (down to $M_{\rm B} = -15$), use a
  higher inclination cutoff ($i = 73^{\circ}$), and a different method
  for selecting cluster members.}. 

A graphical representation of the trend of the bar fraction 
for S0s as a function of environment density 
is shown in Fig.~\ref{envt_comp_fig}. We note that
Fig.~\ref{envt_comp_fig} shows a hint of an increase in the mean bar
fraction toward the dense core of the Coma cluster, however given the
error bars, we cannot say whether this trend is significant.
A comparison of the bar and disk properties, for S0s in Coma and the 
Virgo cluster is discussed in
$\S$~\ref{barredpropdisc}.

We note that to compare to the lowest-density environments (i.e., field galaxies), we
would ideally like to have a comparison sample where bar detection is
done quantitatively via ellipse fits, and where the sample is matched to
ours in both Hubble type (S0 galaxies), luminosity ($M_{\rm V} \lesssim
-18$), and color. However, there is as of yet no field comparison sample that
fulfills all of the above requirements. The two large ellipse fit
bar studies of field galaxies (BJM08 and Aguerri et al.~2009) are not
adequate because they are mismatched in Hubble type and color
(BJM08) or luminosity (Aguerri et al.~2009). Therefore, a 
comparison with these samples could be misleading, in light of recent results showing that the bar fraction 
varies non-monotonically with Hubble type, host galaxy luminosity,
 and color (Nair \& Abraham 2010a). 
A comparison of the bar fraction derived through visual classification to a matched subset 
of S0s from the RC3 is complex, because RC3 galaxies are a mix of field and 
Virgo cluster members. The best candidate for a field 
comparison is the recently released public catalog by Nair \& Abraham 2010b,
containing visual morphologies for $\sim$~14,000 SDSS
galaxies. However, such a study is beyond the scope of this paper
and therefore we defer this comparison to a later work.

Recently M{\'e}ndez-Abreu et al.~(2010) used the Coma Treasury survey
data to analyze the properties of
barred galaxies in the Coma cluster (using visual classification to
detect bars). They do not select a disk sample but look for bars in 
all galaxies (including ellipticals and dwarfs). It is problematic and
unconventional to quote the bar fraction from a sample of disk galaxies
and ellipticals, particularly in the context of studying the bar
fraction as a function of environment, as variations in this bar
fraction can then be caused by the fact that the proportion of
ellipticals to S0s to spirals changes strongly as a function of
environment. For this reason, our study and other studies quote the
bar fraction as the fraction of disk galaxies hosting bars. Comparison
of our work with the results of M{\'e}ndez-Abreu et al.~(2010) is
therefore not straightforward, but nonetheless we attempt a comparison
to check whether our findings are consistent. 
If we include all galaxies from our sample in 
the magnitude range $-23 \le M_{\rm r} \le -14$, to match their 
sample, 
we find a total (visual) bar fraction of 7$\pm$2\% (14/188), while
M{\'e}ndez-Abreu et al.~(2010) find 9\% (secure bars) and 14\% (weak/uncertain bars). 
Although our results are broadly consistent within the uncertainties,
there are several further caveats to this comparison.
M{\'e}ndez-Abreu et al.~(2010) apply the axial ratio constraint $b/a > 0.5$ to their 
whole sample, regardless of morphology, while we only apply an inclination 
cutoff to our bright disk galaxies. In addition to the inclination cutoff, 
we also exclude S0 galaxies deemed to be highly-inclined/edge-on by
eye, since as discussed in $\S$~\ref{inclined}, using only the $i >
60^{\circ}$ cut misses highly-inclined and edge-on S0s with a more circular,
thickened stellar component.  Furthermore, 
in the process of selecting cluster members, M{\'e}ndez-Abreu et al.~(2010) apply 
a color cut where they discard all galaxies that have $g-r$ color greater than 
0.2 above their fit to the cluster red sequence. 

\subsection{Observed and deprojected properties}\label{deproj}
In Figure~\ref{abarebar} we show the observed (solid line) and 
deprojected (dotted line) distributions of the bar semi-major 
axis ($a_{\rm bar}$) and bar ellipticity ($e_{\rm bar}$) for 
the 13 barred S0 galaxies detected through ellipse 
fitting (using both strict and relaxed criteria).
In MJ07, we found that deprojecting the bar semi-major axis and ellipticity
 for a large sample makes only a very small statistical difference (a
 factor of 1.2), on average. However, since our
Coma core sample is small, we deproject the observed radial profiles of ellipticity
 and PA and derive deprojected values of $e_{\rm bar}$ and $a_{\rm bar}$. 
We perform the deprojection using a code developed by  
Laine et al. (2002) and used previously in Laine \etal (2002), 
Jogee \etal (2002a,b), and MJ07. 

We find an observed mean 
bar size of 2.5$\pm$1~kpc (2.9$\pm$1~kpc deprojected; Fig.~\ref{abarPEcomp}) and the mean observed and 
deprojected bar
ellipticity is 0.4$\pm$0.1. 
 It is evident that deprojection does not make a
large difference in the mean $a_{\rm bar}$ and $e_{\rm bar}$. The 
observed and deprojected values of   $a_{\rm bar}$ and $e_{\rm bar}$ of the three bars detected
through the relaxed criteria only are shown as filled and open
circles, respectively in Fig.~\ref{abarebar}. These galaxies satisfy
the relaxed ellipse fit criteria, where
the peak ellipticity of the bar is a local and not a global maximum
due to the combination of factors discussed in
$\S$~\ref{barred}. After deprojection removes the projection effects,
which cause the ellipticity of the disk to be artificially boosted
compared to the bar ellipticity, the peak bar
ellipticity then becomes a global maximum in these  three
galaxies. Thus, after deprojection, these three barred galaxies also pass the
strict ellipse fit criteria. 
The observed and deprojected $a_{\rm bar}$ for these galaxies are 
shorter than average. We note, however, that the bars of the three
galaxies that failed to meet the strict ellipse fit criteria before
deprojection are not necessarily weak bars. The intrinsic ellipticities
$e_{\rm bar}$ of these three bars after deprojection are similar to
the mean value of the whole sample (0.4$\pm$0.1). This can be
understood from the combination of factors discussed in
$\S$~\ref{barred}, particularly the relative orientation of the bar
with respect to the LON and projection effects.  

Fig.~\ref{abarPEcomp} shows that the mean observed and deprojected
bar lengths ($a_{\rm bar}$) in our bright S0 Coma core sample are 
similar to those of S0s in the Virgo cluster (E11).
 The results from 
a KS test are consistent with similar distributions (giving KS ($p = 0.5, D=0.3$) and (0.2, 0.4) for panels (a)
and (b) of Fig~\ref{abarPEcomp}, respectively). We explore the
comparison of the bar and disk properties as a function of environment
density in more detail below.  

\subsection{Properties of disks and bars in the Coma core}\label{barredpropdisc}
We compare our sample of Coma cluster core S0s to 
the properties of S0s in less dense environments, namely to those in
the intermediate-density Abell 901/902 cluster system ($z\sim$~0.165; M09) and
those in the low density Virgo cluster ($z\sim$~0.005; E11).

Figure~\ref{STVbardisk} shows the distributions of (a)~the galaxy 
disk $R_{\rm 25}$ (the isophotal radius
where $\mu_{\rm B}$ reaches 25~mag~arcsec$^{-2}$), (b)~bar semi-major axis $a_{\rm bar}$, measured at the peak 
bar ellipticity $e_{\rm bar}$ for all bars identified 
through ellipse fitting (ER+ES), (c)~the $a_{\rm bar}$/$R_{\rm 25}$ ratio, and 
(d)~peak bar ellipticity $e_{\rm bar}$ for 
the Coma, Abell 901/902, and Virgo (E11) S0 samples. 
$R_{\rm 25}$ values for the E11 Virgo sample are from the RC3. 
$R_{\rm 25}$ for the Coma sample is estimated by ellipse fitting
the galaxies on the ACS F475W images, which approximately 
correspond to SDSS $g$ band. We calibrate the radial profiles 
of surface brightness to mag/arcsec$^{2}$, then convert them 
from $\sim~g$ band (AB) to $B$~mag arcsec$^{-2}$ (Vega) using
Equation~2. For two galaxies, we could not measure $R_{\rm 25}$ radius
because a good fit could not be obtained of the outer disk of the
galaxy due to the presence of a close companion. 
For the M09 Abell 901/902 sample, $R_{\rm 25}$ is calculated 
from the absolute $M_{\rm B}$ magnitudes according to 
\begin{equation}
log(\frac{R_{25}}{kpc}) = -0.249 \times M_{\rm B} -  4.00,
\end{equation}
from Schneider (2006). This formula is derived from an empirical
relation measured for local spirals.  To double-check its validity, we
use it to calculate $R_{\rm 25}$ radii for the Virgo S0s, where we 
already know $R_{\rm 25}$ from RC3. Comparing the calculated values
with those from RC3 confirms that the measured values from RC3 do
follow the above relation, however it under-predicts the true $R_{\rm
  25}$ by $\sim$~1.6~kpc on average. 
All three 
samples have 
similar mean bar and disk properties, but the bar semi-major axis
and disk $R_{\rm 25}$ distributions for Abell 901/902 S0s 
have a tail to larger values. This tail corresponds to the tail 
of brighter S0s present in the Abell 901/902 sample. The mean values of $R_{\rm 25}$ for 
the Coma, Abell 901/902, and Virgo S0s are 6.5$\pm$1.3~kpc, 5.9$\pm$2.8~kpc, 
and 5.0$\pm$2.0~kpc, respectively. The KS statistic reflects these
differences in the Coma, STAGES, and Virgo distributions giving ($p =
0.008$, $D=0.5$) between Coma and Virgo and ($p = 10^{-4}$, $D=0.4$) between Coma and
Abell 901/902.  
The Abell 901/902 and Virgo S0s have similar mean $a_{\rm bar}$/$R_{\rm 25}$ ratios 
of $\sim$~0.4$\pm$0.16, although the range in values is large
($\sim$~0.1--0.9). The KS statistic is consistent with similar 
distributions, giving ($p = 0.9$, $D=0.2$) and (0.6, 0.3) between Coma and Abell
901/902 and Coma and Virgo, respectively.  Coma S0s have a slightly lower mean $a_{\rm bar}$/$R_{\rm 25} =$~
0.35$\pm$0.12. An $a_{\rm bar}$/$R_{\rm 25}$ ratio of $\sim$~0.3$\pm$0.2 has 
also been found for field galaxies averaged over all 
Hubble types (e.g., MJ07, Menendez-Delmestre et al. 2007) and for 
S0 galaxies (Erwin 2005). We note that the range of $a_{\rm
  bar}$/$R_{\rm 25}$ spanned by the three samples is quite large ($\sim$~0.1--0.9), however
although our number statistics are small, this range is
similar
to that found for local field galaxies in MJ07.  

All three samples have very similar distributions in bar 
ellipticity (KS $p \sim$~0.9 and $D\sim$~0.2). The observed bar ellipticities we find for S0s in the Coma cluster as 
well as those for S0s in Virgo and Abell 901/902 are skewed toward
lower values (e.g., mean $e_{\rm bar}\sim$ 0.3--0.4) compared to the bar ellipticities 
in samples dominated by intermediate- to late-type galaxies (e.g.,
MJ07; BJM08; mean $e_{\rm bar}\sim$~0.5--0.7). This difference could 
be intrinsic (i.e., the bars in S0 galaxies are really less elliptical 
than those in later Hubble types), or it could be due to the dilution 
by the bright bulges of the isophotes crossing the end of the bar,
where 
the ellipticity is measured (see
$\S$~\ref{barred}). This effect has been demonstrated by Gadotti (2008).

\subsection{Discussion: implications for the evolution of S0 bars and
 disks as a function of environment density}\label{diskdisc}

What do our results imply for the evolution of bars and disks
in bright S0 galaxies as a function of environment?
We first recapitulate our results.
Using three detection methods (traditional ellipse fit criteria, 
relaxed ellipse fit criteria, and visual classification), 
we found an optical bar fraction of  50$\pm$11\%, 65$\pm$11\%, 
and 60$\pm$11\%, respectively for our 
sample of bright ($M_{\rm V} \lesssim -18$) S0 galaxies
in the central region of the Coma cluster ($\S$~\ref{fbarS0coma}).
 We find that the bar fraction and properties (e.g., $e_{\rm bar}$,
$a_{\rm bar}$)
\textit{in bright S0 galaxies} derived through all 
three of the above methods do not show a statistically significant 
variation (greater than a factor of $\sim$~1.3) 
between the dense central regions of Coma ($n\sim$~10,000 gal/Mpc$^{3}$), 
the intermediate-density Abell 901/902 clusters at $z\sim$~0.165 ($n \sim$~1000
gal/Mpc$^{3}$),
and the low-density Virgo cluster ($n\sim$~300gal/Mpc$^{3}$; Table~\ref{envt_comp}). 
We note that there is a hint that the mean bar fraction 
may show a slight increase as a function of environment density toward the dense
core of the Coma cluster (Fig.~\ref{envt_comp_fig}), however given the
error bars, we cannot say whether this trend is significant. 
Below, we explore what our results may imply for the formation 
and evolution of bars.

It has long been known that DM halo properties 
influence bar formation and evolution. At high 
redshifts (e.g., $z\sim$~5--8), recent theoretical studies of galaxy evolution 
using cosmological initial conditions find that bars 
are triggered by the triaxiality of DM halos and the 
asymmetric DM distribution as a whole (Romano-D{\'{\i}}az et
al.~2008; Heller et al.~2007). These early bars are gas rich, and
quickly decay. Subsequent bar generations form and are destroyed  
during the major-merger epoch (e.g., $z\sim$~2--4)
 due to the rapidly-changing potentials and gas dissipation associated 
with major mergers (Romano-D{\'{\i}}az et
al.~2008). Although DM halos at early times can trigger bar formation 
due to their triaxiality, this triaxiality is diluted 
as disks and other central components
form. The DM halos become more
symmetric, on a timescale that is a function of mass (e.g., Dubinski 1994; Kazantzidis et al.~2004; Heller et
al. 2007). By $z\sim$~1 disks have also become more
massive and stable. Simulations find that large-scale stellar bars
forming at around this epoch are long-lived (Romano-D{\'{\i}}az et
al.~2008; Heller et al.~2007). 
Interestingly, new observational results find that 
the bar fraction for the most massive disks
($M_{*}>10^{11}M_{\odot}$) does not change between $z\sim$~0.6 and 0.2
(Cameron et al.~2010). However, the picture is complicated 
by the fact that for intermediate-mass
disk galaxies ($M_{*}=10^{10.5}$--$10^{11}M_{\odot}$), the bar
fraction builds up 
by a factor of two over that redshift range. In addition, 
at $z\sim$~0, the bar fraction and properties are a non-monotonic
function of the host galaxy properties, such as stellar mass, 
luminosity, color, Hubble type, and SF history (BJM08; M09; Aguerri et al.~2009;
Barazza et al.~2009; Weinzirl et al.~2009; Laurikainen et al.~2009;
Gadotti 2011; Nair \& Abraham 2010a).

The picture above does not directly discuss environmental 
effects. In fact, there are still few theoretical and observational 
studies addressing this aspect of bar evolution. 
However, increasingly the emerging picture is suggesting that the
frequency and properties of bars do not appear to be a sensitive
function of environment (van den Bergh 2002; Aguerri et al. 2009; M09; 
Barazza et al.~2009; Cameron et al.~2010; although see Giuricin et al. 1993 and Elmegreen et al. 1990).

How do the above results make sense in light of many 
theoretical studies that show that galaxy interactions 
can trigger bars in unbarred galaxies (e.g., Noguchi 1988; 
Mihos \& Hernquist 1996)? We present a tentative picture below, 
considering the competing effects present in galaxy clusters. 
If a disk galaxy is  sufficiently dynamically cold (i.e., Toomre $Q \lesssim 1.5$), 
it is susceptible
to non-axisymmetric $m=2$ instabilities (e.g., bars) whether spontaneously
induced (e.g., Toomre 1981; Binney \& Tremaine 1987) or 
tidally induced (e.g., Noguchi 1988; Hernquist 1989; Heller \& 
Shlosman 1994; Mihos \& Hernquist 1996; Jogee 2006 and 
references therein). The effect of the interaction
depends on the geometry (i.e., prograde or retrograde encounter),
with retrograde encounters having little to no effect on an already
existing bar (e.g., Gerin, Combes, \& Athanassoula 1990;
Steinmetz \& Navarro 2002; Romano-D{\'{\i}}az et al.~2008;
Aguerri \&  Gonz{\'a}lez-Garc{\'{\i}}a 2009). 

At $z < 1.5$, as clusters assemble and field galaxies fall into
the existing cluster potential, let us now ask how the fraction
and properties of bars in S0s  might
be expected to differ from the field environment.
In a rich cluster, where the projected galaxy number density ($n$) and galaxy velocity
dispersion ($\sigma$) is high, the timescale for close 
interactions (or collision timescale, $t_{\rm coll}$) will be short.
We can estimate this timescale using:
\begin{equation}
t_{\rm coll} = \frac{1}{n\sigma_{\rm gal}A},
\end{equation}
where $n$ is the galaxy number density, $\sigma_{\rm gal}$ is the
galaxy velocity dispersion, and $A$ is the cross-section for close interactions
defined as
\begin{equation}
A = \pi f(2r_{\rm gal})^{2}.
\end{equation}
We assume $f$ is unity, $r_{gal}\sim$~10~kpc.
For the Coma core  $\sigma_{\rm
gal}\sim$~900~km/s, and $n\sim$~10,000 gal/Mpc$^{3}$ (The \& White 1986)
giving a short timescale for close interactions $t_{\rm coll}\sim$~90~Myr.

However, although these close galaxy-galaxy interactions are frequent
in a rich cluster, the large
galaxy velocity dispersions present mean that each single encounter
will be a {\it high speed one}. Unlike single slow, strong encounters,
a single high-speed encounter will typically not induce a large amount of
tidal damage and not lead to major mergers.  As a result, three factors
may make it difficult for new bars to be induced in disk galaxies in a
cluster. Firstly, single high-speed encounters may not be as effective in
inducing bars as slow, strong encounters, because the timescale over
which gravitational torques act is short.
Secondly, over time, the cumulative  effect of many  high-speed 
and weak encounters (galaxy harassment), can tidally
heat disks (e.g., Moore et al. 1996; Aguerri \& Gonz{\'a}lez-Garc{\'{\i}}a
2009), making such disks dynamically hot (with Toomre $Q > 1.5$), and thus less susceptible to
bar instabilities.  Finally, in a cluster environment, the
accelerated star formation history (e.g., Balogh et al. 2004, Blanton
et al.~2005b; Hogg et al. 2003) 
as well as physical processes
such as ram pressure stripping  (Gunn \& Gott 1972; Larson et al.~1980;
Quilis et al.~2000) will make S0 disks gas-poor, thus making them
less bar-unstable. We therefore speculate  that these three factors,
namely the predominance of
high speed encounters over slow ones, the tidal heating of S0 disks,
and the low gas content of S0s in rich clusters, make it difficult
for many new bars to be induced in S0 disks as they infall from
a field-like  to a cluster-like environment.
This  scenario may explain, at least in part, our findings that
there is no strong variation in the optical bar fraction  of S0s
across the range of low density to high density environments
characterized by Coma, Virgo, and Abell 901/902 in our study, as
well as claims  by other studies that there is no difference in the 
bar fraction between
clusters and the field (van den Bergh 2002, Aguerri et al. 2009,
Barazza et al.~2009; M09). Our interpretation for the result that the
bar fraction is not greatly enhanced in the dense Coma cluster is also in 
agreement with that of M{\'e}ndez-Abreu et al. (2010). 

We note that it is possible that in rich clusters, the
above effects, particularly the tidal heating, may  cause existing
bars to weaken. However, this effect is hard to robustly demonstrate
observationally as  the measured bar ellipticity is diluted by
relatively large bulges in S0s (which dominate the disk population
in clusters),  while in the field, the  disk population is dominated
by spirals where such a dilution is not as severe (see $\S$~\ref{barredpropdisc}).

It is also important to note that the arguments above, which explain
why the bar fraction might not be greatly enhanced in rich clusters
compared to the field,  would lead to a rather different prediction
for how the bar fraction in groups would compare to that in the field.
In a group, the
number density is moderately high ($n \sim$~10)  but the galaxy velocity
dispersions are typically  low ($\sigma \sim$~100; Tago et al.~2008). Therefore slow, strong
encounters are expected to be frequent in groups. Such encounters are
likely to induce  extra bars in disk galaxies compared to the field,
particularly given the fact that the disks will not be stripped of their
cold gas in groups as they would in rich clusters.  
In this context, we note that indeed
higher bar fractions have been reported for early-type galaxies in
binary pairs (Elmegreen et al.~1990) and early-type galaxies that are
disturbed/interacting (Varela et al. 2004).

\section{Bars and disk features in Coma dwarfs}\label{dwarfs}

In addition to investigating bars in high-mass galaxies, we 
also take advantage of the exquisite resolution of the ACS
($\sim$~50~pc at the distance of Coma) to 
search for bars and other disk features (e.g., spiral arms, edge-on disks) in the
numerous dwarf galaxies in the central regions of the Coma cluster. Are some of these galaxies
the remnants of late-type spirals that have gone through processing in
a dense cluster environment? In Virgo, some early-type dwarfs are
known to host features (e.g., lenses, bars, spiral arms; Sandage \& Binggeli 1984; Binggeli \&
Cameron 1991; Jerjen et
al. 2000; Barazza et al. 2002; Lisker et al.~2006; Lisker et al.~2007;
Lisker \& Fuchs 2009) suggesting the presence of a disk. Not only can
such features provide clues to the formation history of these systems,
but the presence or absence of bar structures has implications
for the conditions necessary for bar formation and growth in
galaxies (e.g., M{\'e}ndez-Abreu et al. 2010). 

\subsection{Identifying dwarf galaxies}\label{dwarfget}
As outlined in $\S$~\ref{data}, we use a magnitude cut
at $M_{\rm I (814)} = -18.5$ (AB mag), (roughly corresponding to $M_{\rm V} = -18$ Vega mag)
to separate dwarf and normal galaxies.  A montage of some of 
the faint, low-mass dwarfs in our Coma core sample is shown in Fig.~\ref{dwarfmont}. 
Fig.~\ref{dwarfmag}a shows the distribution of absolute
magnitude $M_{\rm  I(814)}$ of the galaxies in the faint 
dwarf sample. Fig.~\ref{dwarfmag}b shows where the dwarf galaxies lie on a plot of the
$\mu_{\rm e}$ (the surface brightness at $R_{\rm e}$) vs. absolute
magnitude $M_{\rm I(814)}$. Effective radii and $\mu_{\rm e}$ 
are from Hoyos et al.~(2010), derived
through single-component S\'ersic fits. 

Prior to applying the unsharp masking technique ($\S$~\ref{dwarfmethod})
 on the dwarf sample, we pick out good
candidates through a cut in surface brightness ($\mu_{\rm e} < 25$~mag~arcsec$^{-2}$) and radius ($R_{\rm 90} > 100$~pc). 
These cuts remove very low surface-brightness objects, and those where we are unlikely to resolve disk 
structure. We choose a size cut at $R_{\rm 90} > 100$~pc, which is twice the ACS PSF at the distance of Coma. 
Out of the 417 dwarf galaxies, we
find 333 dwarfs that satisfy these criteria. The $\mu_{\rm e}$ vs. 
$M_{\rm I (814)}$ distribution of these galaxies is plotted with yellow
triangles in Fig.~\ref{dwarfmag}b.



\subsection{Identifying bars and other disk features in dwarfs}\label{dwarfmethod}
 
In many dwarf galaxies
disk features may not be readily apparent by eye (or traditional
quantitative methods such as ellipse fitting) because their
amplitude is very low and is overwhelmed by the smooth light from the galaxy. Which method is
sensitive enough to detect faint spiral/bar structure in such systems?
Jerjen et
al. (2000) used residuals from subtracting the azimuthally-averaged
light profile of the galaxy from the original image to discover hidden spiral
features in IC~3328 with deep VLT observations. They
analyzed the spiral structure using Fourier expansion, finding that
the amplitude of the spiral is only $\sim$~3 to 4\%. However, upon
further analysis of a larger sample of 20 Virgo dwarfs, Barazza et
al. (2002) find that some spiral or bar-resembling residuals may be
artifacts from the combination of the increasing ellipticity and
twisting isophotes (due to triaxiality) present
in these galaxies and not actual spiral structure. Fourier
decomposition is similarly unsuccessful in many galaxies. Barazza et
al. (2002) find that a
much better method seems to be unsharp masking (e.g., Schweizer \& Ford 1985;
Mendez et al.~1989; Buta \& Crocker 1993; Colbert et al.~2001; Erwin \& Sparke 2003). In this method, no 
assumptions about the light profile/inclination of the galaxy
are necessary.  
Recently, Lisker et al.~(2006) also successfully  employed unsharp
masking on $\sim$~470 Virgo dwarfs to look for evidence of bar/spiral
structure. 
Graham et al.~(2003) discovered two dwarf galaxies with spiral structure in the Coma cluster
using unsharp masking as well as subtracting a symmetrical model to 
reveal non-symmetrical disk features (one of these galaxies
is COMAi125937.988p28003.56 in Table~\ref{dwarfinfo}, while the 
other is not covered by the Coma ACS Treasury survey). Chilingarian et al.~(2008)
use unsharp masking to find disk features in dwarf galaxies in 
Abell 496. 
We therefore use the unsharp masking method to seek out bar (or
spiral) structures in the Coma cluster core dwarf sample. 

We perform unsharp masking for the 333 dwarf galaxies that fit the
criteria outlined in $\S$~\ref{dwarfget}. 
First, we smooth the
galaxy images by convolving with a Gaussian using the IRAF task
$GAUSS$. Then we divide the original galaxy image by the
smoothed image. We choose the Gaussian smoothing kernel size to be
$\sim$~25 pixels, corresponding to $\sim$~625~pc for our galaxies. We
also try a range of smoothing lengths from $\sim$~15--45 pixels
($\sim$~375--1125~pc) for a
subsample of the galaxies and find no substantial change in the
results. A point made by Lisker et al.~(2006) is that in some cases,
it is desirable to use an elliptical smoothing aperture matched to the
outer ellipticity and PA of the galaxy, in order to avoid spurious
detections that resemble an edge-on disk. For this reason, in all
cases where we suspect that the galaxy host an inclined/edge-on disk,
we also perform the unsharp masking using an elliptical PSF to ensure
that the structures found are not spurious detections. 

We find bars and/or spiral arms in 13 galaxies out of the 
333 dwarfs in the unsharp-masked subsample. An additional eight galaxies show evidence
of an inclined disk (or ambiguous inclined disk or bar). The galaxies where we find structure are 
listed in Table~\ref{dwarfinfo}.
Fig.~\ref{usmask} shows examples of the residuals due to different types of structures:
(a)~spiral arms only, (b)~bar and spiral arms, (c)~inclined disk, 
(d)~bar and/or spiral, (e)~ambiguous bar or inclined disk embedded in a stellar 
halo. The galaxies with disk structure are overplotted as cyan 
points in Fig.~\ref{dwarfmag}. Most (76\%) of the galaxies
where we find disk structure are brighter than $M_{\rm I(814)} = -16$ (AB mag). 
We discuss the possible implications of
these results below. 

\subsection{Discussion: barred dwarf galaxies in the Coma core}\label{dwarfdisc}

Using visual inspection and unsharp masking we find only 13 galaxies
with bars and/or spiral arms in our Coma core dwarf subsample of 333
galaxies with $\mu_{\rm e} < 25$~mag~arcsec$^{-2}$ and $R_{\rm 90} > 100$~pc. Does this result imply that
faint/dwarf galaxies with disks are very rare 
 within the Coma population, or rather that any
existing disks in these galaxies are too dynamically hot to be
unstable to disk instabilities? 

Studies have long been finding early-type dwarf galaxies with
spiral/bar structure in
Virgo and Fornax (e.g., Sandage \& Binggeli 1984; Binggeli \&
Cameron 1991; Jerjen et al.~2000; Barazza et al.~2002; Lisker et al.~2006; Lisker et
al.~2007; Lisker \& Fuchs
2009). Lisker et al.~(2006) search through 476 Virgo early-type dwarfs
and find unambiguous stellar disk structure (bar/spiral) in 14 of them, while
another 27 have probable or possible disk features. Some 
authors have speculated that anywhere from 5\% to 50\% of Virgo
early-type dwarfs have disk structure, depending on the magnitude range under
scrutiny (Lisker et al.~2006; Lisker \& Fuchs 2009). 

Approaching the search for disks in early-type dwarfs a different way, 
 Aguerri et al.~(2005)
investigate a sample of galaxies in Coma with $-18 \le M_{\rm B} \le
-16$ and classify them into two types dE or dS0 depending on their
surface brightness
profile. Galaxies whose surface brightness profiles are well fitted by a single S\'ersic
law are classified as dEs, and those with surface brightness profiles fitted with a
S\'ersic plus exponential profile are classified as dS0s. 
Aguerri et al.~(2005) find that about 30\% of their Coma dwarf sample
cannot be fitted well by a single S\'ersic law, suggesting that
 early-type dwarfs with disks may not be scarce in
Coma. Graham \& Guzman (2003) found evidence for outer disks in three out of
a sample of 18 Coma early-type dwarfs, modeling the surface
brightness profiles using a Sersic function in combination with either a central point source or a 
resolved central Gaussian component using high-resolution $HST$
images. 

While it is still unclear whether all early-type dwarfs with disk
structure represent a distinct class of galaxies 
that are the product of a single formation mechanism, one plausible
scenario is that they are formed through processing of faint, late-type
spirals and irregulars in cluster environments (e.g., Kormendy 1985; Lin \&
Faber 1983; Graham et al. 2003; Lisker et al. 2006). 
This processing includes the loss of their gas
through ram-pressure stripping (e.g., Gunn \& Gott 1972) 
as well as the cumulative effects of harassment in a
cluster environment (e.g., Moore et al.~1996). 
The simulations of Mastropietro et al.~(2005) have shown that indeed
late-type spirals can be reprocessed into early-type dwarfs through
cluster processes (such as harassment), and that these dwarfs do
retain their stellar disk structure. On the other hand, observations
of a small number of 
 isolated early-type dwarf galaxies (e.g., Fuse et al.~2008;
 Hern{\'a}ndez-Toledo et al.~2010)  argue against 
cluster transformation processes as the sole explanation for the
formation of these objects. 


We find evidence
of only 13 dwarfs hosting disk instabilities (bar and/or spiral arms)
 in our unsharp-masked subsample of 333
dwarfs. This result is in broad agreement with the 
findings of M{\'e}ndez-Abreu et al. (2010), who find a paucity of barred disks for 
Coma galaxies fainter than $M_{r} = -17$. As also suggested by M{\'e}ndez-Abreu et al. (2010),
 these results imply that although 
it is possible that as many as $\sim$~30\% of dwarf galaxies in Coma may have a
disk component (Aguerri et al.~2005), the majority do not have the necessary conditions 
to form or maintain  bar and
spiral instabilities, namely a disk that is dynamically cold.
This result is consistent with previous studies, showing a paucity of
\textit{thin} disks in lower-luminosity dwarf galaxies (S{\'a}nchez-Janssen et
al. 2010; Yoachim \& Dalcanton 2006).

\section{Summary}\label{summary}

We use ACS F814W images from the \textit{Hubble Space Telescope} ACS Treasury survey of the Coma cluster
at $z\sim$~0.02 to study the fraction and properties of barred galaxies in the central 
region of Coma, 
the densest environment in the nearby Universe. The available data span 274 arcmin$^{2}$, where
 approximately 75\% of the data are within 0.5~Mpc of the
 cluster center, and contain thousands of
 sources down to a limiting magnitude of $I =$~26.8 mag in F814W (AB
 mag).  We initially select 469 cluster 
members and split the sample with a magnitude cut at 
$M_{\rm V}\lesssim -18$ (Vega mag). 
Using this magnitude cut, we investigate two different regimes: (1)~the
fraction and properties of bright $M_{\rm V} \lesssim -18$ S0 
galaxies and (2)~the presence of bars and other disk features (e.g.,
bars and spiral arms) in faint/dwarf galaxies in the Coma
core. Our results for the two populations are described below. 

(1)~For S0 galaxies: We select a sample
of 32 bright S0 galaxies based on visual classification
supplemented by multi-component decompositions in ambiguous cases
($\S$~\ref{nondwarfget}). After discarding 12 highly-inclined galaxies, 
we identify and
characterize bars in the remaining 20 moderately-inclined S0s 
using three methods: ellipse fits where the bar 
is detected through strict criteria (the peak bar ellipticity 
$e_{\rm bar}$ is required to be a global maximum in the radial profile of 
ellipticity); ellipse fits where 
the bar is detected through relaxed criteria (which do not require the peak bar ellipticity $e_{\rm bar}$ 
to be a global maximum); and visual 
classification. We find:
\begin{itemize}
\item{The optical bar fraction for our 
bright S0 sample is: 50$\pm$~11\%, 65$\pm$~11\%, and 60$\pm$~11\%
 based on ellipse fits with traditional and relaxed
criteria, and visual classification, respectively (Table~\ref{bar_stats}).}
\item{We compare to results from studies in less dense environments
  (Abell 901/902 and Virgo) and find that the bar fraction, as well 
as the mean quantitative properties of the S0 bars and disks (e.g., $R_{25}$, 
$a_{\rm bar}$, $e_{\rm bar}$) do not show a statistically significant
variation, within the error bars, for samples of matched S0s in environment densities ranging 
from $n\sim$~300~gal/Mpc$^{3}$ (Virgo), $n\sim$~1000~gal/Mpc$^{3}$ (Abell 901/902), and $n\sim$~10,000~gal/Mpc$^{3}$ (Coma), 
with high galaxy velocity dispersions $\sigma \sim$~800~km/s (Table~\ref{envt_comp}, Fig.~\ref{STVbardisk}). 
We note that there is a hint that the mean bar fraction 
may show a slight increase as a function of environment density toward the dense
core of the Coma cluster (Fig.~\ref{envt_comp_fig}), however given the
error bars, we cannot say whether this trend is significant. 
We speculate that the bar fraction  among S0s is not dramatically
enhanced in rich  clusters compared to low density environments
due to several factors. Firstly, S0s in rich clusters  are likely 
to be more stable to bar instabilities because they are dynamically 
heated by the cumulative effect of many  high-speed,  weak encounters 
(galaxy harassment), and additionally are gas poor as a result of  
ram pressure stripping and  accelerated star formation. Secondly, 
individual high-speed encounters in  rich clusters may be less effective 
than individual slow strong encounters in inducing bars. The combination 
of these effects precludes an enhancement in the bar fraction for S0 galaxies in cluster environments compared to the field. 
Our results are in agreement 
with recent observational studies which find no difference in the fraction 
of barred galaxies with environment density over all Hubble types.}
\end{itemize}

(2)~For faint/dwarf galaxies: We select a sample of 
417 galaxies fainter than $M_{\rm I (814)} = -18.5$  (AB mag; $\S$~\ref{dwarfget})
where  we utilize our $\sim$~50~pc
resolution to look for disk structures such as bars and spiral arms
using visual classification of unsharp-masked images. After applying unsharp 
masking to a subsample of 333 dwarfs  
($\mu_{\rm e} < 25$~mag~arcsec$^{-2}$, $R_{\rm 90} > 100$~pc; $\S$~\ref{dwarfmethod}),
 we find only 13 dwarf galaxies with a
 bar and/or spiral arms, and an additional eight galaxies where an inclined disk may be present 
(Fig.~\ref{usmask}). These results suggest that either disks
are not common in these galaxies in the Coma cluster core, or that any
disks present are too hot to form instabilities. \\

I.M., S. J,. and T. W. acknowledge support from the National Aeronautics 
and Space Administration (NASA) LTSA grant NAG5-13063, NSF grant AST-0607748,
and $HST$ grants GO-11082 and GO-10861 from STScI, which is operated by 
AURA, Inc., for NASA, under NAS5-26555. SJ thanks the Excellence Cluster Origin and Structure of
the Universe in Garching, Germany, sponsored by  TUM, the
 Ludwig-Maximilians-Universitt (LMU),  the Max-Planck-Institutes,
 and the European Southern Observatory (ESO). P.E. was supported by DFG 
Priority Program 1177 (`Witnesses of Cosmic
History:  Formation and evolution of black holes, galaxies and their
environment'). DC acknowledges support from the UK Science and 
Technology Facilities Council under grant
ST/H002391/1.




{}                            


\setcounter{table}{0}
\begin{deluxetable}{lccccccccc}
\tabletypesize{\scriptsize}
\tablewidth{0pt}
\tablecaption{Optically barred bright ($M_{\rm V} \lesssim -18$) S0s.}
\tablehead{
\colhead {Bar ID}&
\colhead {Galaxy ID } &
\colhead {RA}&
\colhead {DEC}& 
\colhead {$M_{\rm I (814)}$}&
\colhead {Bar detection method}&
\colhead {$e_{\rm bar}$}&
\colhead {$e_{\rm bar}$}&
\colhead {$a_{\rm bar}$}&
\colhead {$a_{\rm bar}$}\\
\colhead {} &
\colhead {} &
\colhead {(AB mag)}&
\colhead {}&
\colhead {}&
\colhead {}&
\colhead {obs}&
\colhead {dep}&
\colhead {obs (kpc)}&
\colhead {dep (kpc)}\\
\colhead {(1)} &
\colhead {(2)} &
\colhead {(3)}&
\colhead {(4)}&
\colhead {(5)}&
\colhead {(6)}&
\colhead {(7)}&
\colhead {(8)}&
\colhead {(9)}&
\colhead {(10)}\\
}
\startdata
B1&COMAi125710.767p272417.44  &      194.29486  &        27.404846 &  -20.1 &  ES, ER, V & 0.35 & 0.26 & 2.97 & 2.99\\
B2&COMAi125833.136p272151.77  &      194.63806  &        27.364383 &  -20.0  &  ES, ER, V & 0.68 & 0.54 & 3.19 & 3.52\\
B3&COMAi125928.728p28225.90  &       194.86970  &        28.040531 &   -19.7  &  ES, ER & 0.49 & 0.40 & 2.26 & 2.44\\
B4&COMAi125929.956p275723.26  &      194.87481  &        27.956462 &   -21.1  &  ES, ER, V & 0.47 & 0.51 & 4.36 & 5.04\\
B5&COMAi125946.775p275825.88  &      194.94490  &        27.973856 &   -20.6  &  ES, ER, V & 0.39 & 0.70 & 2.22 & 2.82\\
B6&COMAi125956.707p275548.62  &      194.98628  &        27.930173 &   -19.9   &  ES, ER, V & 0.55 & 0.46 & 3.42 & 4.00\\
B7&COMAi13022.156p28249.08  &        195.09231  &        28.046968 &   -20.6       &  ES, ER, V & 0.37 & 0.46 & 1.92 & 2.23\\
B8&COMAi13038.731p28052.22  &        195.16137  &        28.014507 &  -20.3    &  ES, ER, V & 0.47 & 0.31 & 3.57 & 3.77\\
B9&COMAi13042.753p275816.88  &       195.17814  &        27.971355 &  -21.3   &  ES, ER, V & 0.25& 0.31 & 1.46 & 1.55\\
B10&COMAi13042.833p275746.98  &       195.17846  &        27.963052&    -20.3&  ES, ER, V & 0.43 & 0.46& 2.48&  2.64\\
B11&COMAi125930.825p275303.42  &      194.87843  &        27.884283 & -21.1 &  ER, V & 0.33& 0.35 & 1.99 & 2.33\\
B12&COMAi13017.020p28350.10  &        195.07092  &        28.063917 &   -19.0 &  ER, V & 0.29 & 0.40 & 1.14 & 1.45\\
B13&COMAi13027.971p275721.54  &       195.11654  &        27.955985 &   -20.1  &  ER, V & 0.31 & 0.47 & 2.09 & 2.79\\
\enddata
\tablecomments{(1)~Bar ID; (2)~Galaxy ID as given in the Coma Treasury survey DR2 (Hammer et al. 2010); (3)~RA (J2000); 
(4)~DEC (J2000); (5)~$M_{\rm I (814)}$ absolute magnitude in AB mag; (6)~Bar detection method: `ES' - strict ellipse
fit criteria ($\S$~\ref{barred}), `ER' - relaxed ellipse fit criteria
($\S$~\ref{barred}), and `V' - visual classification on direct image ($\S$~\ref{visual}; (7)~Observed peak
bar ellipticity $e_{\rm bar}$; (8)~Deprojected peak bar ellipticity; (9)~Observed bar semi-major axis $a_{\rm bar}$ measured 
at $e_{\rm bar}$; (10)~Deprojected bar semi-major axis ($\S$~\ref{barredpropdisc}). 
\label{barinfo}
}
\end{deluxetable}

\setcounter{table}{1}
\begin{deluxetable}{lccccc}
\tabletypesize{\scriptsize}
\tablewidth{0pt}
\tablecaption{Optical bar fraction for bright ($M_{\rm V} \lesssim -18$) S0s based on different methods.}
\tablehead{
\colhead {Method } &
\colhead {Unbarred}&
\colhead {Barred}& 
\colhead {$f_{\rm bar,opt}$}\\
}
\startdata
Ellipse fit, strict & 10& 10 & 50$\pm$11\%\\
Ellipse fit, relaxed     & 7 & 13 & 65$\pm$11\%\\
Visual classification    & 8 & 12 & 60$\pm$11\%\\
\enddata
\tablecomments{optical bar fraction for the 20 moderately inclined ($i<60^{\circ}$) bright
($M_{\rm V} \lesssim -18$) S0 galaxies. Barred galaxies are characterized
through: (1)~ellipse fitting using the strict criteria (where $e_{\rm bar}$ is 
required to be a global maximum in the ellipticity profile), (2)~ellipse fitting 
using relaxed criteria ($e_{\rm bar}$ can be a local maximum), and (3)~visual classification. 
\label{bar_stats}
}
\end{deluxetable}


\setcounter{table}{2}
\begin{deluxetable}{lcccc}
\tabletypesize{\scriptsize}
\tablewidth{0pt}
\tablecaption{Optical bar fraction for bright ($M_{\rm V} \lesssim -18$)
S0s in  different environments.}
\tablehead{
\colhead {Study } &
\colhead {Environment}&
\colhead {Number density (gal/Mpc$^{3}$)}&
\colhead {Velocity dispersion (km/s)}&
\colhead {S0 $f_{\rm bar}$ for $M_{\rm V} \lesssim -18$}\\
\colhead{(1)}&
\colhead{(2)}&
\colhead{(3)}&
\colhead{(4)}&
\colhead{(5)}\\
}
\startdata
 & & \\
\multicolumn {4}{c} {\small Bars identified through ellipse fitting (strict criteria; $\S$~\ref{barred})} \\
\hline\\
this work & Coma central region, $z\sim$~0.02 & 10,000 & 900$^{(a)}$ & 50$\pm$11\%  (10/20)\\
M09$^{(b)}$  & Abell 901/902 clusters, $z\sim$~0.165 & 1000 & 800--1200$^{(c)}$ &39$\pm$5\% (38/98)\\
E11  & Virgo,  $z\sim$~0 &  300 & 400--750$^{(d)}$ & 44$\pm$14\% (12/27) \\
 & & \\
\multicolumn {4}{c} {\small Bars identified through ellipse fitting (relaxed criteria; $\S$~\ref{barred})} \\
\hline\\
this work & Coma central region $z\sim$~0.02 & 10,000 & 900 &65$\pm$11\%  (13/20)\\
M09$^{(b)}$ & Abell 901/902 clusters, $z\sim$~0.165 & 1000 & 800--1200 &48$\pm$5\% (47/98)\\
E11  & Virgo, $z\sim$~0 & 300 & 400--750 &48$\pm$14\% (13/27) \\
 & & \\
\multicolumn {4}{c} {\small Bars identified through visual classification} \\
\hline \\
this work & Coma central region, $z\sim$~0.02 & 10,000 & 900 &60$\pm$11\% (12/20) \\
T81$^{(e)}$  & Coma central region, $z\sim$~0.02 & 10,000 &900 &42$\pm$7\%  (19/45)\\
M09$^{(b)}$  & Abell 901/902 clusters, $z\sim$~0.165 & 1000 & 800--1200 &55$\pm$5\% (54/98)\\
E11$^{(f)}$  & Virgo, $z\sim$~0 & 300 & 400--750 &59$\pm$9\%  (16/27)\\
\enddata
\tablecomments{
T81: Thompson 1981; M09: Marinova et al.~2009; E11: Erwin et al.~(in prep.)\\
(a)~The \& White~(1986)\\
(b)~We use a sub-sample from M09, with the criteria outlined in $\S$~\ref{s0comp}.\\
(c)~Heiderman et al.~(2009)\\
(d)~Binggeli et al.~(1987)\\
(e)~Bar classification is performed on ground-based KPNO plates.\\
(f)~For this paper, visual classification is performed on the E11 sample
by P.E., I.M., and S.J. using the criteria outlined in $\S$~\ref{visual}.\\
\label{envt_comp}
}
\end{deluxetable}

\setcounter{table}{3}
\begin{deluxetable}{lccccc}
\tabletypesize{\scriptsize}
\tablewidth{0pt}
\tablecaption{Galaxies in the faint ($M_{\rm V} > -18$) sample
where we find disk structure through unsharp masking.}
\tablehead{
\colhead {Galaxy ID } &
\colhead {RA}&
\colhead {DEC}& 
\colhead {Visit}&
\colhead {$M_{\rm I (814)}$}&
\colhead {Structure}\\
\colhead {(1)} &
\colhead {(2)} &
\colhead {(3)}&
\colhead {(4)}&
\colhead {(5)}&
\colhead {(6)}\\
}
\startdata
COMAi13030.949p28630.18  &       195.12895  &        28.10838  &   02   &-17.6 &  inclined disk\\
COMAi13041.192p28242.38  &       195.17163  &        28.04510  &   08   &-17.8  &  spiral \\
COMAi13018.883p28033.55  &       195.07867  &        28.00931  &   09   &-18.1 &  bar \\
COMAi13013.398p28311.81  &       195.05583  &        28.05328  &   10   &-17.7 &  inclined disk \\
COMAi13007.727p28051.91  &       195.03219  &        28.01441  &   10   &-13.3 &  inclined disk\\
COMAi125930.062p28237.71  &       194.87525  &        28.04380  &   13  & -13.4 &  bar/inclined disk\\
COMAi125904.792p28301.21  &       194.76997  &        28.05033  &   14  & -18.3  &  bar\\
COMAi125911.545p28033.38  &       194.79811  &        28.00927  &   14  & -18.2  &  inclined disk\\
COMAi125953.930p275813.76  &       194.97471  &        27.97048  &   18 &  -16.9  &  spiral\\
COMAi125937.988p28003.56  &       194.90827  &        28.00098  &   19  & -18.0  &  bar + spiral\\
COMAi13035.418p275634.05  &       195.14758  &        27.94279  &   22  & -17.7 &  bar\\
COMAi13024.823p275535.89  &       195.10342 &        27.92663  &   23   & -18.2 &  bar + spiral\\
COMAi125950.181p275445.54  &       194.95909  &        27.91265  &   25 &  -17.5 &  bar\\
COMAi125820.533p272546.03  &       194.58555  &        27.42945  &   45 &  -18.3 &  spiral\\
COMAi125815.275p272752.96  &       194.56364  &        27.46471  &   45 &  -17.2 &  inclined disk\\
COMAi125814.969p272744.81  &       194.56237  &        27.46244  &   45 &  -15.5  &  bar + spiral\\
COMAi125825.308p271200.04  &       194.60545  &        27.20001  &   59 &  -18.4 &  bar + spiral\\
COMAi125623.788p271402.30  &       194.09912  &        27.23397  &   63 &  -18.1  &  bar + spiral\\
COMAi125638.099p271304.09  &       194.15875  &        27.21780  &   63 &  -16.4  &  bar\\
COMAi125845.297p274650.75  &       194.68873  &        27.78076  &   75 &  -15.3  &  bar/inclined disk\\
COMAi125845.906p274655.90  &       194.69126  &        27.78219  &   75 &  -14.7 &  bar/inclined disk\\
\enddata
\tablecomments{(1)~Galaxy ID as given in the Coma Treasury survey DR2 (Hammer et al. 2010); (2)~RA (J2000); 
(3)~DEC (J2000); (4)~$HST$ visit number; (5)~$M_{\rm I (814)}$ absolute magnitude in AB mag; (6)~type of disk structure
detected through unsharp masking (see $\S$~\ref{dwarfmethod}). 
\label{dwarfinfo}
}
\end{deluxetable}



\clearpage
\setcounter{figure}{0}
\begin{figure}
\epsscale{0.6}
\plotone{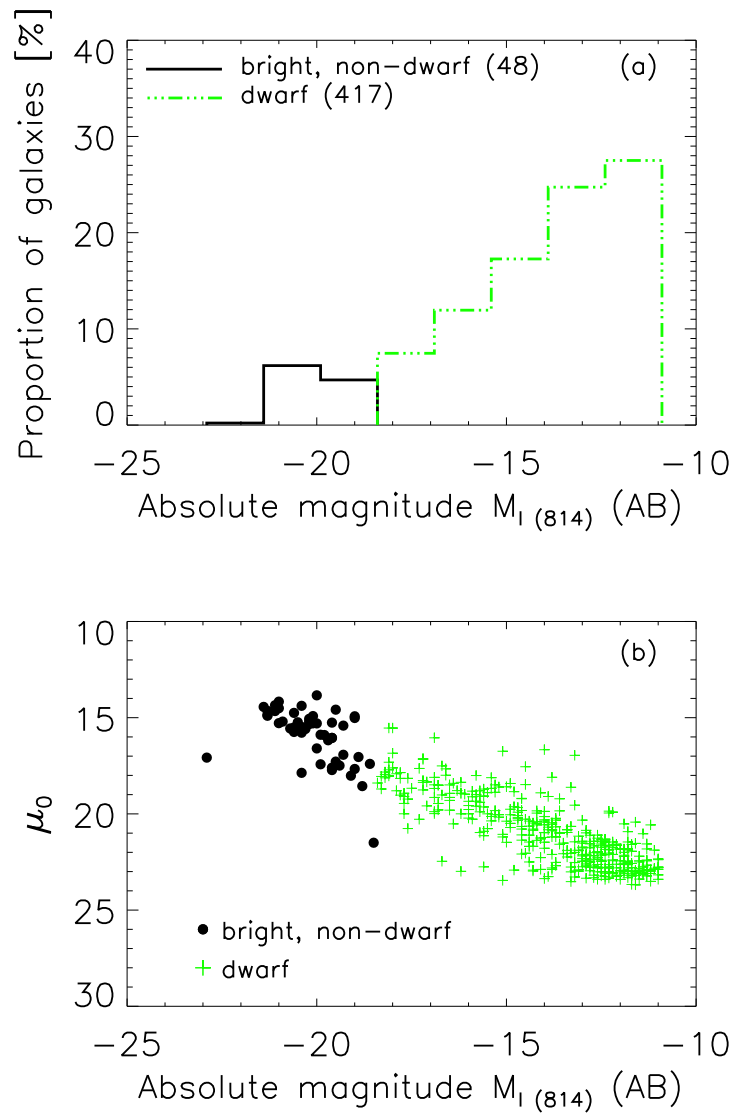}
\caption{\textbf{(a)}~Absolute magnitude ($M_{\rm I (814)}$)
  distribution of the bright, non-dwarf (black solid line; $M_{\rm V} \lesssim -18$) and
dwarf (green dash-dot line; $M_{\rm V} > -18$) galaxies in our Coma core cluster member sample
  ($\S$~\ref{data}). Most galaxies are dwarfs with $M_{\rm V}
  > -18$.  \textbf{(b)}~Central surface brightness $\mu_{\rm 0}$ (from the $SExtractor$
source catalogs in DR2) vs. 
absolute magnitude $M_{\rm I (814)}$ for the bright, non-dwarf (black circles) and dwarf (green plus)
cluster core samples. 
\label{totpropbf}}
\end{figure}

\clearpage
\setcounter{figure}{1}
\begin{figure}
\epsscale{0.9}
\plotone{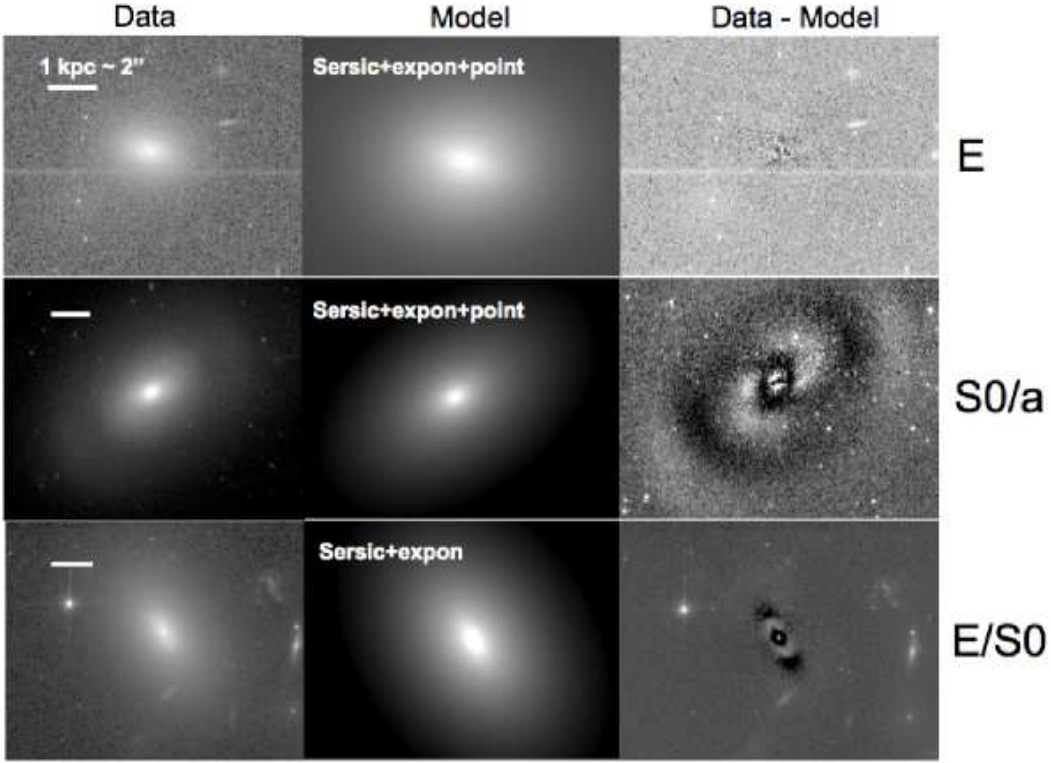}
\caption{
For the 10 visually-ambiguous E/S0 galaxies, single and multiple 
component decompositions were performed with GALFIT on the 
2D light distribution to help determine whether
they are an elliptical (E) or S0. This figure shows examples of the 
data (left), GALFIT model (middle), and (data-model) residuals (right) 
for three of these galaxies, which ended up being classified after 
decomposition  ($\S$~\ref{nondwarfget}) as E (top row), S0/a 
(middle row), and still  ambiguous E/S0 (bottom row). 
For the galaxy (COMAi125930.268p28115.17) in the top row 
the best fit model consists of a  S\'ersic component of half light 
radius $R_e \sim 1.4$~kpc, within which is embedded a compact
disk and a nuclear point source.  This galaxy does not make the
cut to be an S0 as it  lacks a bulge and extended outer disk, and
it is not included in the analysis of the bar fraction for S0s.
The galaxy in the second row (COMAi125938.323p275913.84) is 
classified as an S0/a because after fitting a  (S\'ersic+Exponential) 
model, corresponding to a bulge and extended outer disk, one can
see extended spiral  structure  in the residuals.  Since spiral 
structures only  exist in disks, this confirms the presence of an outer
disk.  We classify this galaxy as an S0/a rather than an Sa because 
the spiral structure is revealed in the residuals and is not  readily 
visible on the direct image. The galaxy (COMAi125950.103p275529.47) 
in the third row  is classified as an ambiguous E/S0 because, based on
the residuals and other factors ($\chi^{2}$ and fit parameters), it
is still not possible to determine whether this galaxy is more likely 
to be an elliptical with an inner debris disk or an S0.    
\label{bdmontage}}
\end{figure}

\clearpage
\setcounter{figure}{2}
\begin{figure}
\epsscale{0.7}
\plotone{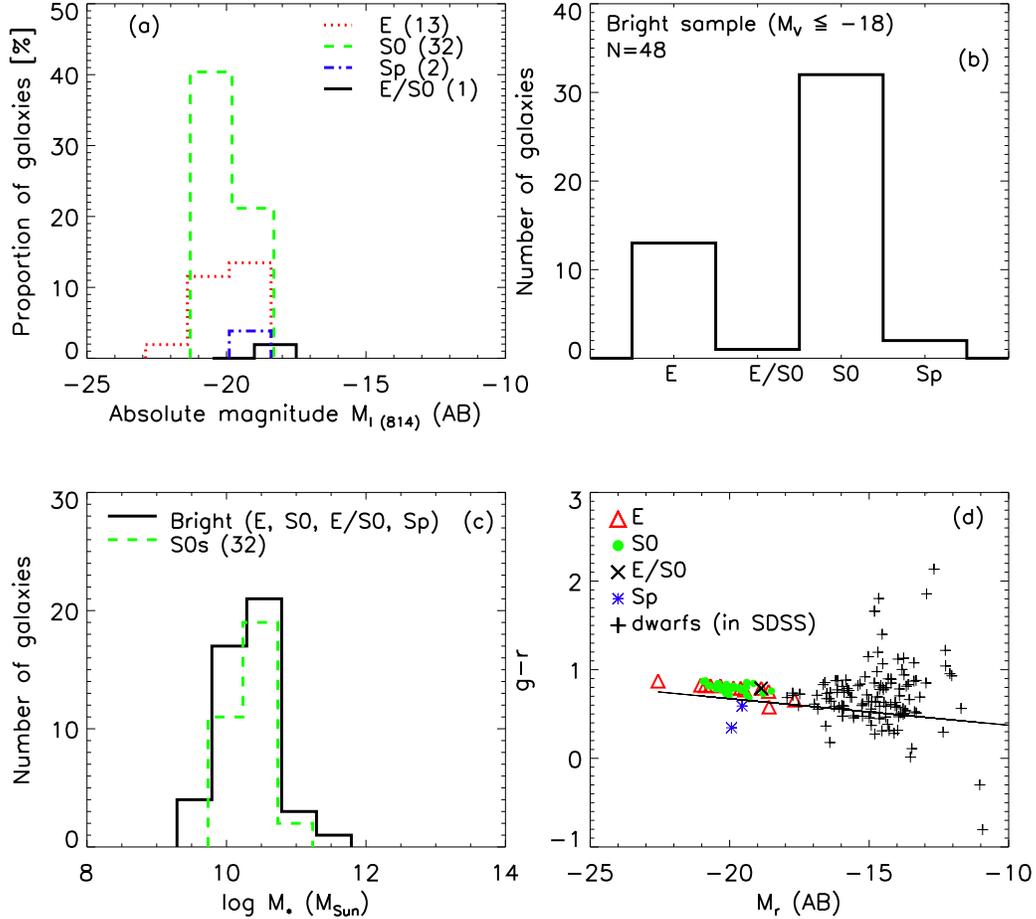}
\caption{\textbf{(a)}~Absolute magnitude
  distribution of bright ($M_{\rm V} \le -18$) galaxies in our Coma core sample
  ($\S$~\ref{nondwarfget}).  \textbf{(b)}~Distribution of morphological types (E, E/S0, S0, Sp) in the bright non-dwarf sample. 
S0 galaxies (32), comprise 94\% of the bright disk galaxies. Morphological types
are from visual classification, supplemented with 2D, multi-component decomposition
for visually ambiguous cases ($\S$~\ref{nondwarfget}). \textbf{(c)}~Stellar mass distribution of the bright galaxies 
in our Coma core sample, with the S0 galaxies shown in green. The 
S0s have masses between 10$^{9.5}$ and 10$^{11} M_{\odot}$. \textbf{(d)}~$g-r$ color-magnitude
diagram of the bright cluster sample and a subset (30\%) of the dwarf
  sample with available SDSS magnitudes. We overplot the relation from
  Blanton et al.~(2005) for the break between the red sequence 
and blue cloud. Most elliptical and S0 galaxies lie on the red sequence. 
\label{totprop}}
\end{figure}

\clearpage
\setcounter{figure}{3}
\begin{figure}
\epsscale{1}
\plotone{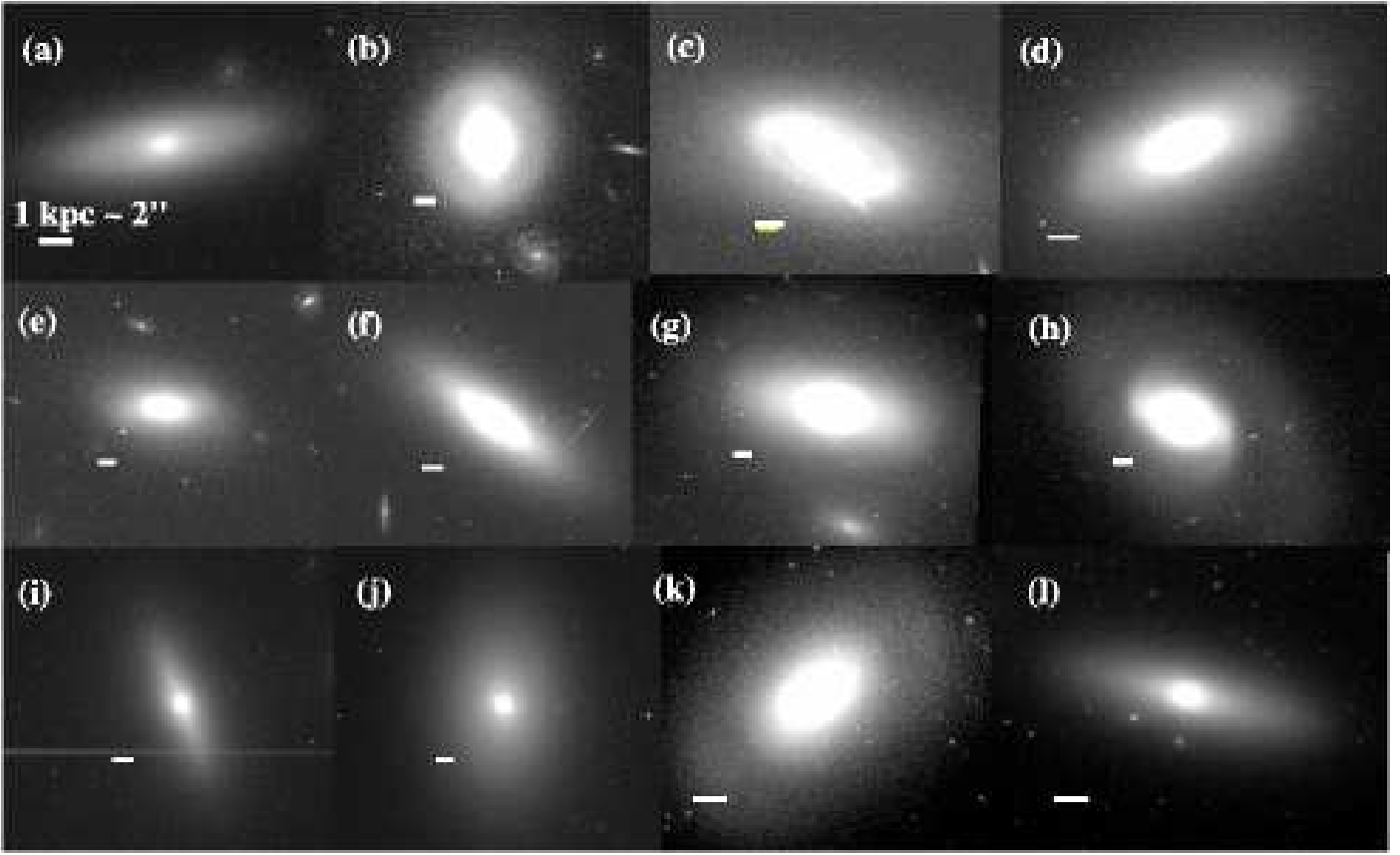}
\caption{Examples of some S0 galaxies from our 
bright sample ($\S$~\ref{nondwarfget}). The scale bars show 1~kpc. The
galaxies shown are:~(a)~COMAi125704.336p273133.26,
(b)~COMAi125710.767p272417.44, (c)~COMAi125833.136p272151.77,
(d)~COMAi125832.060p272722.85, (e)~COMAi125928.728p28225.90,
(f)~COMAi125929.404p275100.51, (g)~COMAi125929.956p275723.26,
(h)~COMAi125930.825p275303.42, (i)~COMAi125931.455p28247.62, 
(j)~COMAi125932.789p275900.95, (k)~COMAi125938.323p275913.84, 
(l)~COMAi125939.657p275713.86. 
\label{s0montage}}
\end{figure}

\clearpage
\setcounter{figure}{4}
\begin{figure}
\epsscale{1}
\plotone{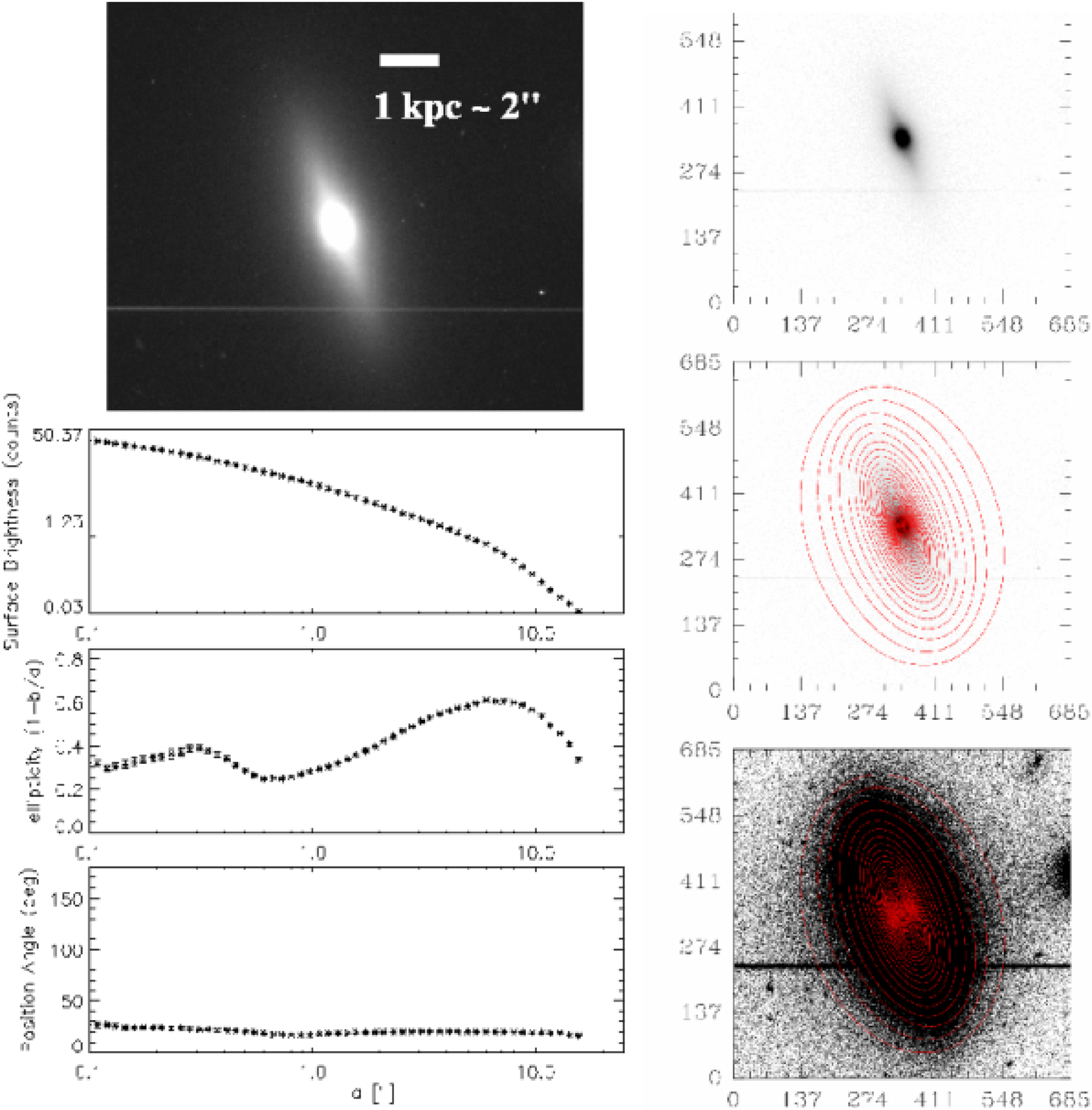}
\caption{Left: $HST$ image and radial profiles of surface brightness, $e$, and PA
  of a highly-inclined S0 with an outer, diffuse, thickened stellar
component (see $\S$~\ref{inclined}). Right: the ellipse fits are overlaid onto the galaxy
  image. The top two panels are shown with a stretch that enhances the
  thin disk and boxy bulge, while the bottom panel shows the outer
  disk. The thickened, diffuse, outer stellar component causes the 
outermost isophotes to have $e \sim$~0.4, which is less 
than the quantitative inclination cut of $e > 0.5$. Therefore, we classify this 
galaxy as highly-inclined using visual classification according to the 
criteria outlined in $\S$~\ref{inclined}. 
\label{visincexample}}
\end{figure}

\clearpage
\setcounter{figure}{5}
\begin{figure}
\epsscale{1}
\plotone{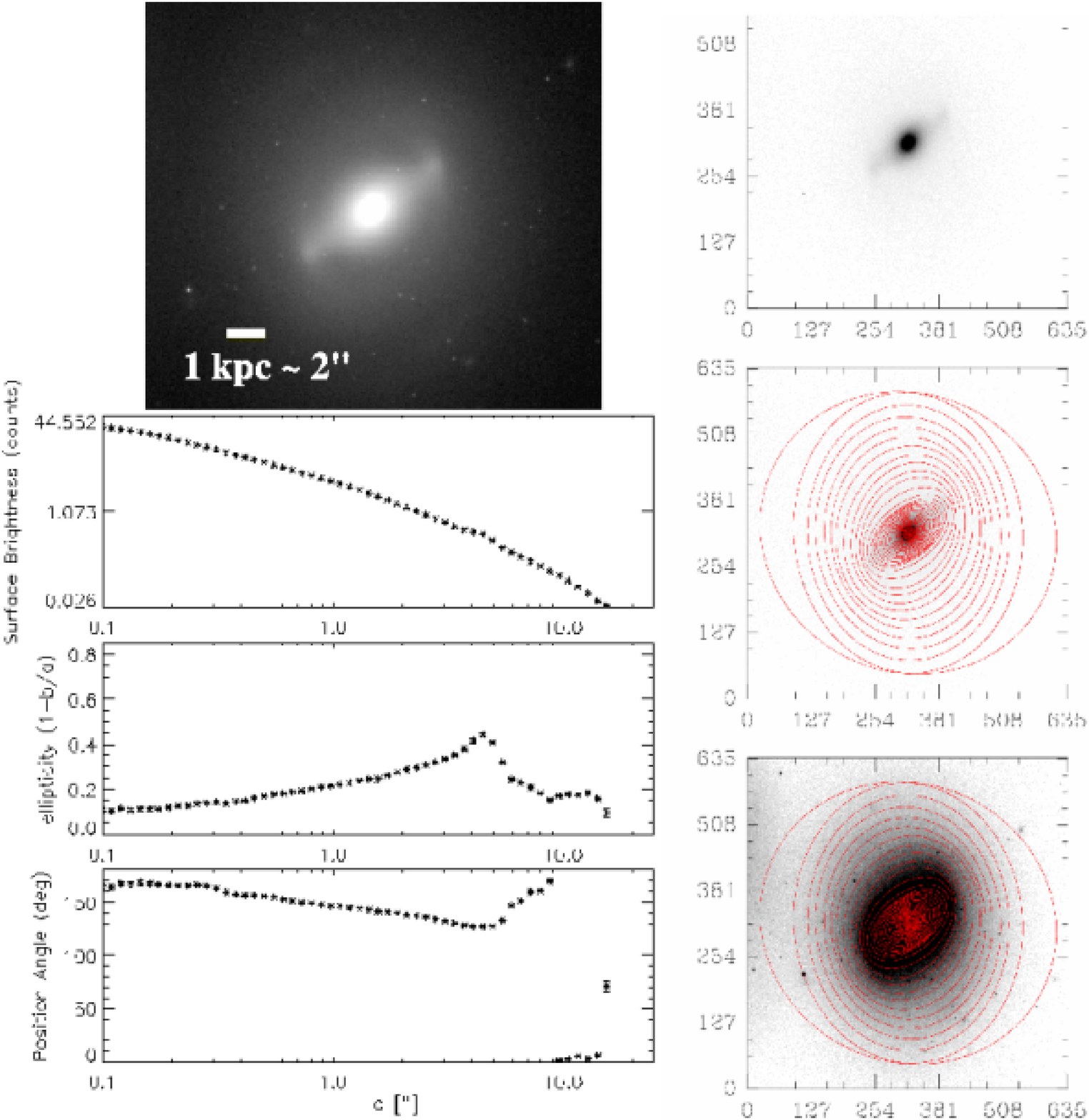}
\caption{Left: $HST$ image and radial profiles of surface brightness, $e$, and PA
  of a barred cluster galaxy. In this example, the traditional bar signature is
  evident in the smooth rise of the $e$ to a global maximum of $\sim$~0.4, while the
  PA remains relatively constant in the bar region. 
The $e$ then drops and the PA changes, indicating the transition to
  the disk region. Right: the ellipse fits are overlaid onto the galaxy
  image. The top two panels are shown with a stretch that enhances the
  inner disk and bar regions, while the bottom panel shows the outer
  disk. See $\S$~\ref{barred} for details.  
 \label{ellipseex1}}
\end{figure}

\clearpage
\setcounter{figure}{6}
\begin{figure}
\epsscale{1}
\plotone{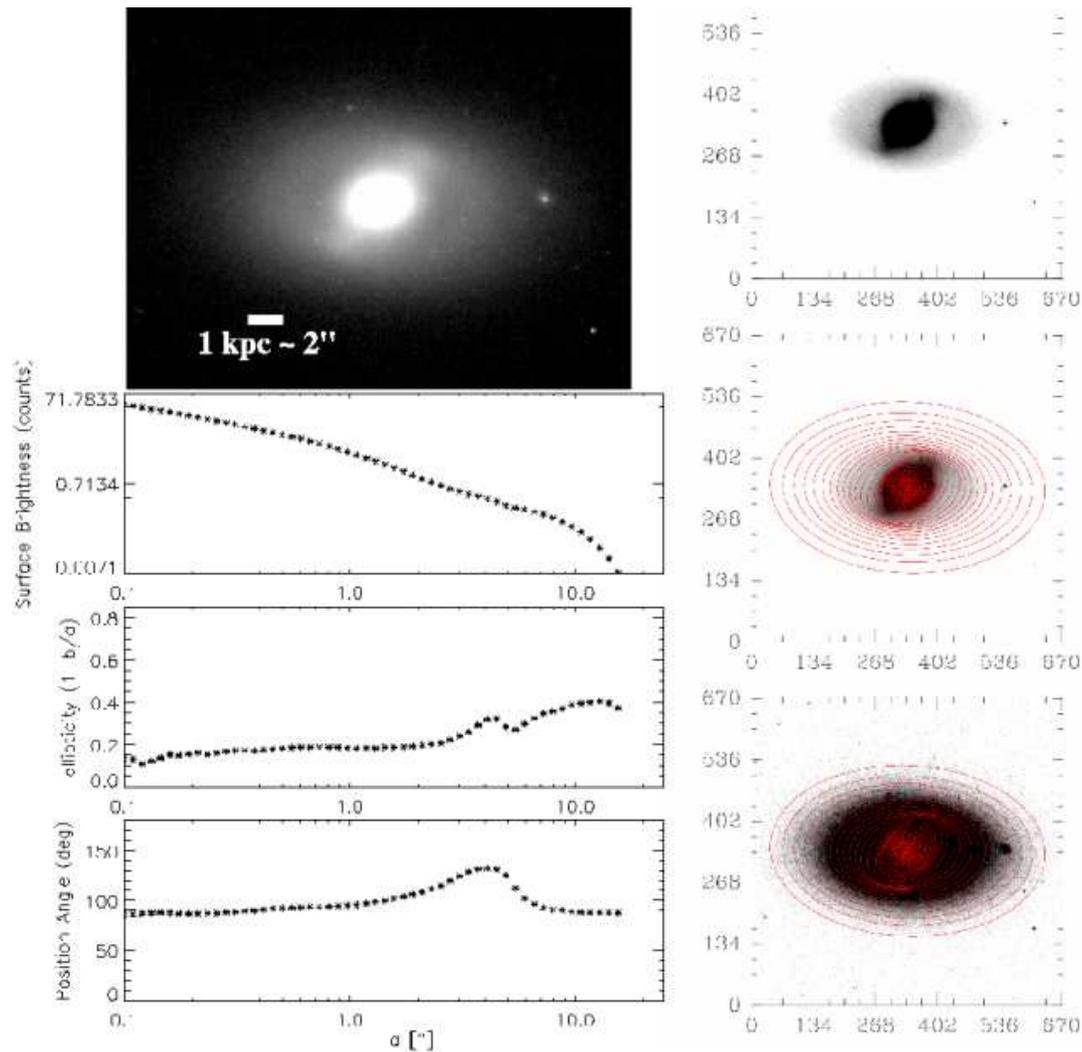}
\caption{
The panels are as in Fig.~\ref{ellipseex1}, but here 
we show an example of a barred galaxy that does not meet
the strict ellipse-fit criterion requiring that $e_{\rm bar}$
is the {\it global} maximum in the $e$ radial profile. In 
this case, the observed outer disk ellipticity $e_{\rm disk}$
is higher than $e_{\rm bar}$, making it a {\it local}
maximum in the $e$ radial profile. 
This happens due to a combination of properties of
the galaxy: (1)~the galaxy is inclined ($i\sim~51^{\circ}$) causing the outer
disk to  be elongated along the line of nodes with a significant
ellipticity ($e_{\rm disk} = 0.37$); (2)~the stellar bar is significantly
offset (by $\sim$~45$^{\circ}$)  with respect to the line of nodes 
and hence its intrinsic axial ratio is diluted by projection effects; 
(3)~the stellar bar has a significant
fraction of its length inside a very luminous bulge, and 
the measured bar ellipticity is diluted to lower 
values than the true $e_{\rm bar}$.
Therefore this galaxy is identified as `barred' through the relaxed 
ellipse-fitting criteria. We find three such cases among the bright 
Coma S0 galaxies (see $\S$~\ref{barred}).
 \label{ellipseex2}}
\end{figure}

\clearpage
\setcounter{figure}{7}
\begin{figure}
\epsscale{0.9}
\plotone{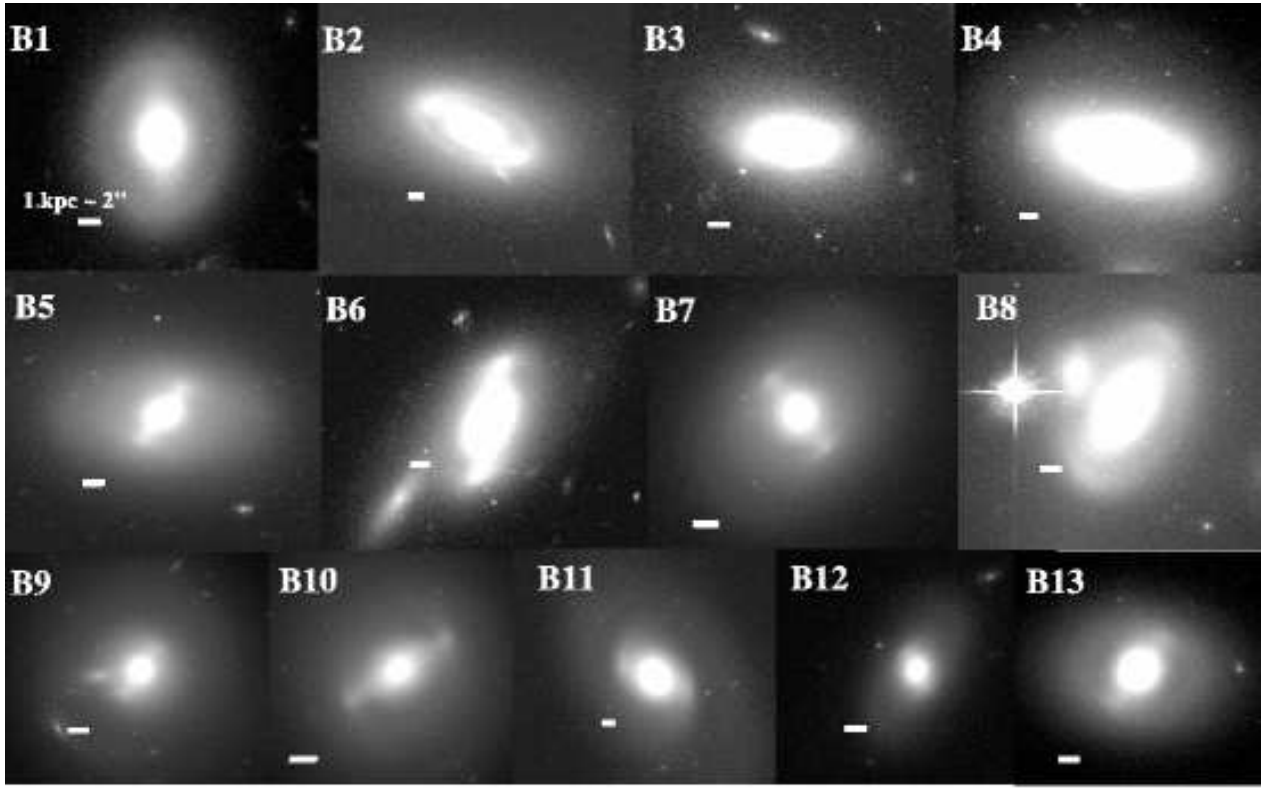}
\caption{Barred bright ($M_{\rm V} \lesssim -18$) S0 galaxies in the Coma cluster core found 
through ellipse fits (strict and relaxed criteria) and visual classification 
(see $\S$~\ref{diskmethod} and Table~\ref{barinfo}). All bars
identified through the relaxed ellipse fit criteria are also
identified by visual classification and vice versa. Bright stars such
as the one in B8 are masked during the fitting.  
 \label{barstamps}}
\end{figure}

\clearpage
\setcounter{figure}{8}
\begin{figure}
\epsscale{0.7}
\plotone{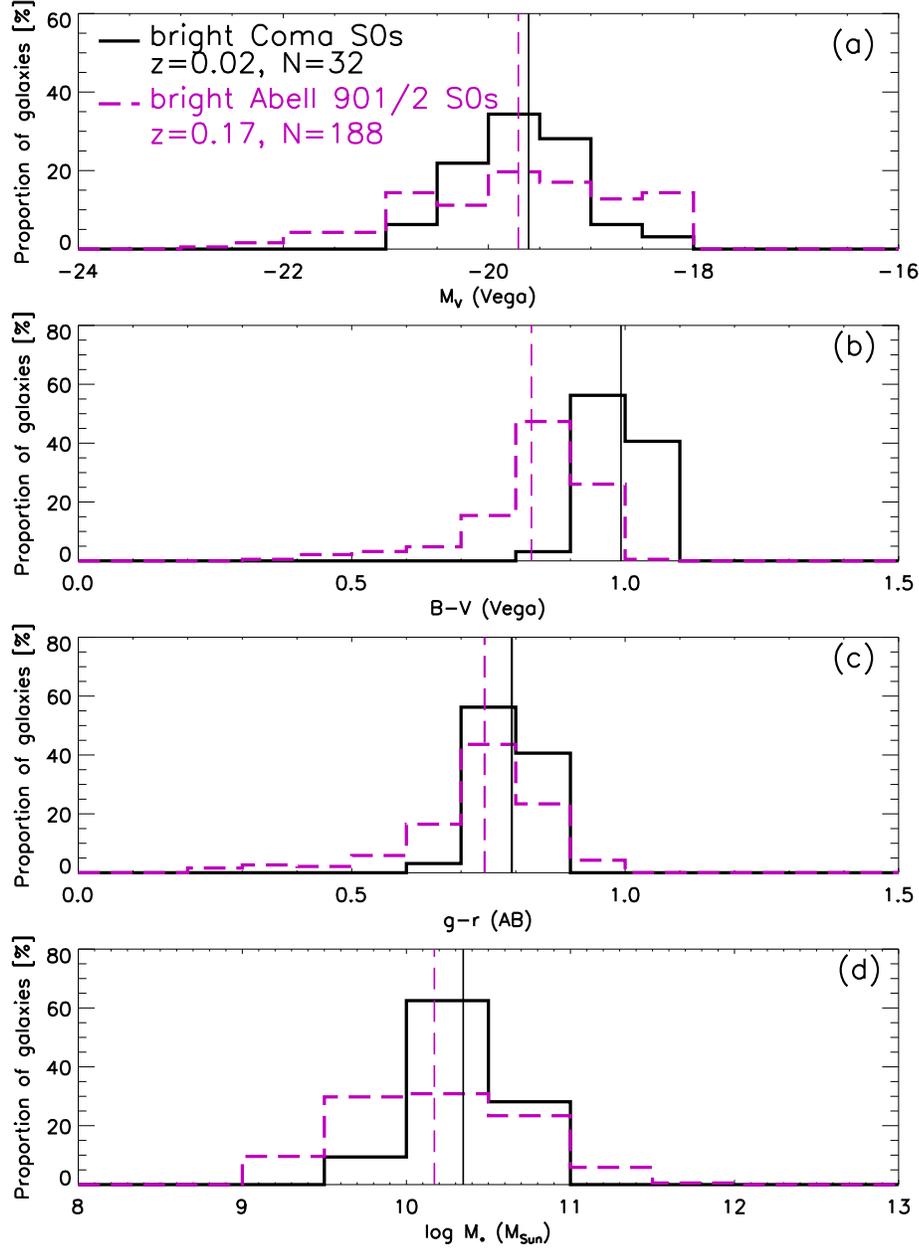}
\vskip 1 in 
\caption{The properties of bright S0 galaxies in 
the Abell 901/902 cluster system (dashed pink lines) and 
Coma (solid black line). The vertical lines show the mean
values for each distribution. 
The two samples are well-matched in 
mean luminosity, $g-r$ color, and stellar mass. However, the results 
of a Kolmogorov-Smirnov (KS) test show differences in the overall distributions for 
$g-r$ color and stellar mass (KS ($p = 0.002$, $D=0.4$) and (0.005,0.3), respectively)
and are inconclusive for the distributions of $M_{\rm V}$ (KS ($p = 0.13$, $D=0.2$)). This is
likely due to the fact that the Abell 901/902
sample has a tail of galaxies with masses both lower and higher than 
the Coma core sample, translating, respectively, into a 
tail of bluer colors and brighter absolute magnitudes. The Abell 901/902 S0s
appear $\sim$~0.2 mag bluer in $B-V$ color, on average (KS ($p = 6*10^{-16}$, $D=0.8$)). We 
believe that this $B-V$ color offset  is not  real
and is  caused by the fact that the color transformations derived 
by Jester et al. (2005)  for stars may not be  adequate for S0s. 
 \label{m09sampprops}}
\end{figure}

\clearpage
\setcounter{figure}{9}
\begin{figure}
\epsscale{0.7}
\plotone{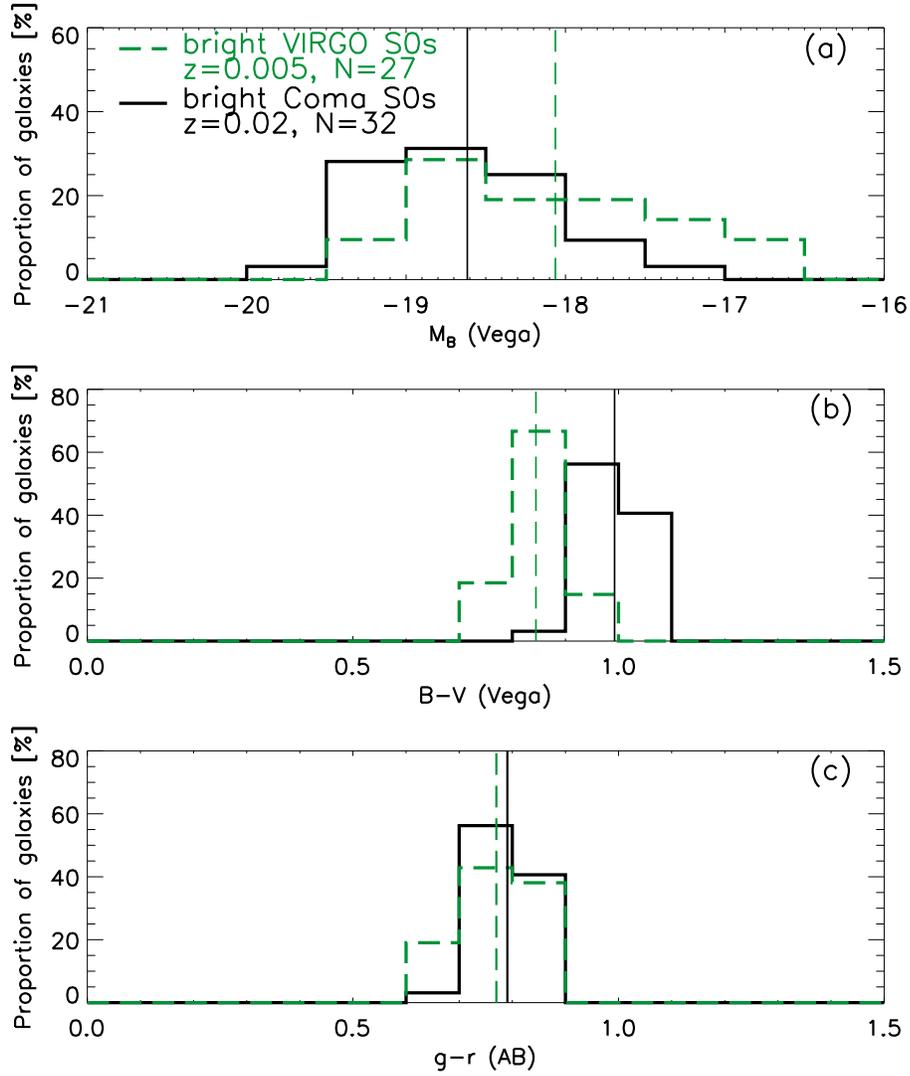}
\vskip 1 in 
\caption{The properties of bright S0 galaxies in 
the Virgo cluster (dashed green line) and 
Coma (solid black line). The vertical lines show the mean
values for each distribution. The two samples are well-matched in 
mean luminosity and $g-r$ color. A KS test shows differences in 
the overall distribution of $M_{\rm B}$ absolute magnitude ($p = 0.02$, $D=0.4$), 
whereas the distributions of $g-r$ color are similar (KS $p = 0.3$, $D=0.3$). 
While the Virgo S0s appear bluer in $B-V$ ($\sim$~0.15 mag on 
average, with KS ($p = 8 * 10^{-11}$, $D = 0.9$)) we believe that this color offset 
is not  real since the measured $g-r$ colors (from SDSS in panel (c)) 
for  the Coma and Virgo  samples agree well.  A possible reason 
for the offset in $B-V$ is that the  the color transformations derived 
by Jester et al. (2005)  for stars may not be  adequate for S0s.
 \label{pe11sampprops}}
\end{figure}

\clearpage
\setcounter{figure}{10}
\begin{figure}
\epsscale{0.7}
\plotone{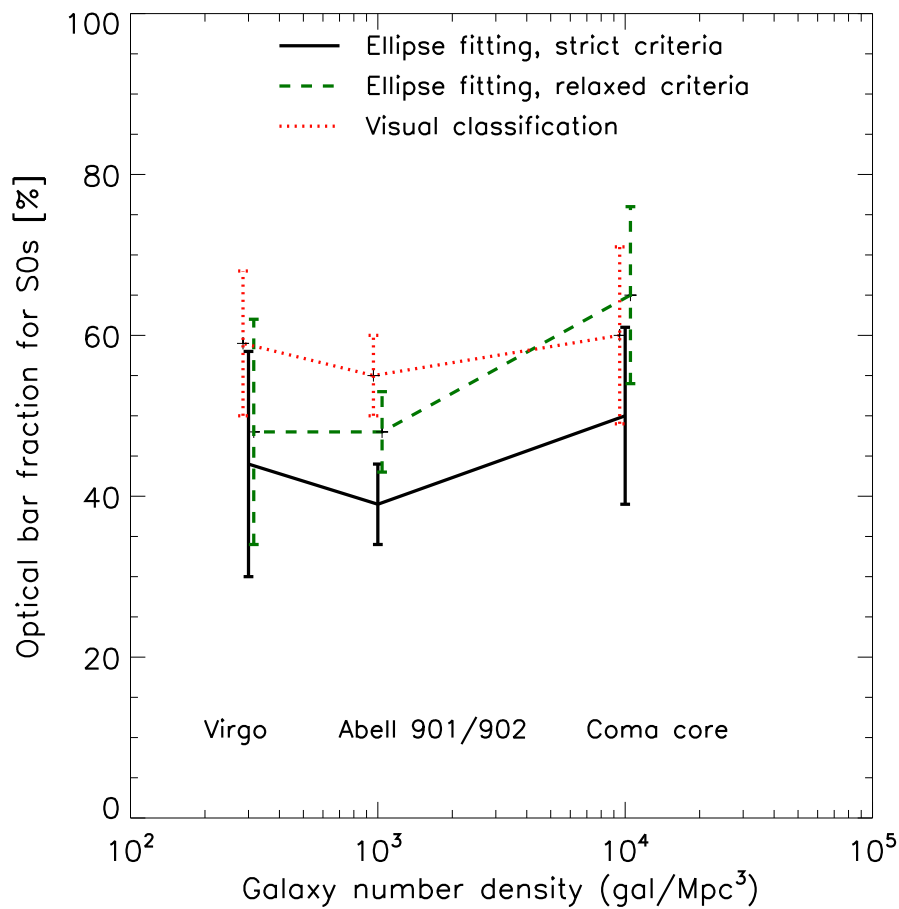}
\caption{The optical bar fraction for S0 galaxies characterized
  through three methods (ellipse fitting with strict criteria, ellipse
  fitting with relaxed criteria, and visual classification) as a
  function of environment density. The different environments probed
  are the high-density core of Coma ($n\sim$~10,000~gal/Mpc$^{3}$), the
  intermediate-density Abell
  901/902 cluster system ($n\sim$~1000~gal/Mpc$^{3}$), and the
  low-density Virgo
  cluster ($n\sim$~300~gal/Mpc$^{3}$; $\S$~\ref{s0comp}). The bar
  fraction for S0s does not show a statistically significant variation
  across the environments probed, within the error bars. 
 \label{envt_comp_fig}}
\end{figure}


\clearpage
\setcounter{figure}{11}
\begin{figure}
\epsscale{0.7}
\plotone{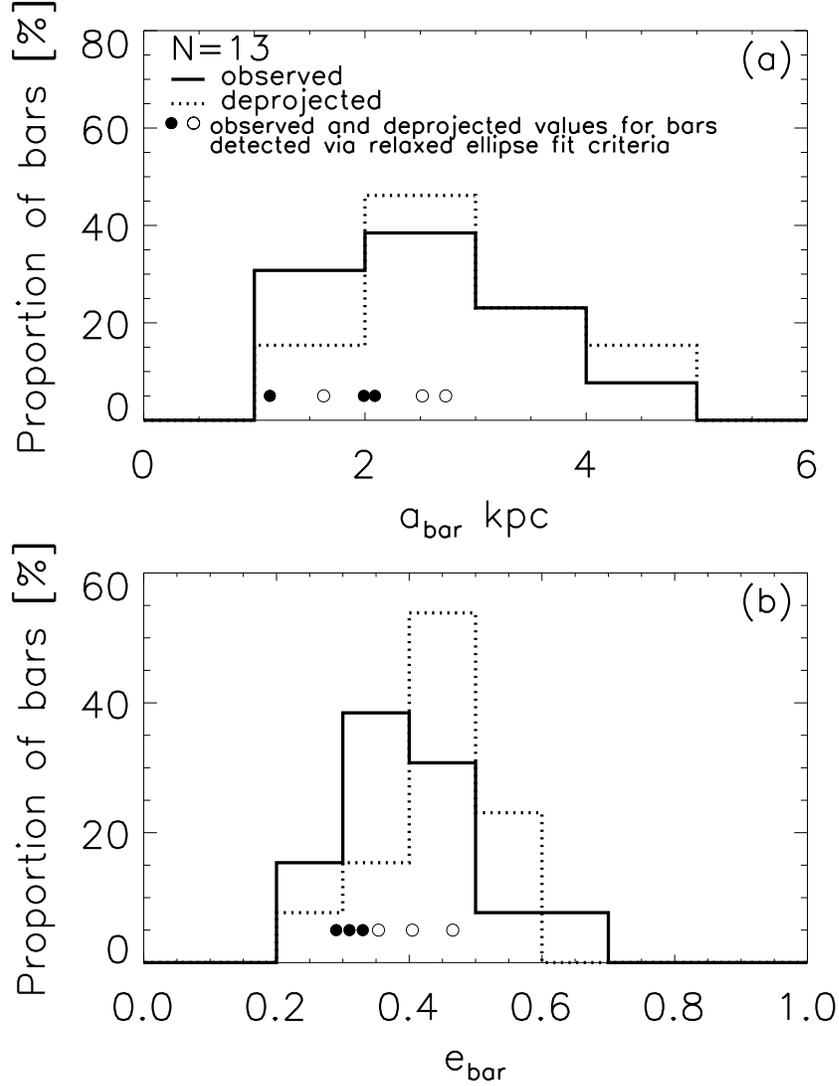}
\vskip 1 in 
\caption{~Observed (solid line) and deprojected (dotted line)
 bar size ($a_{\rm bar}$) \textbf{(a)} and ellipticity $e_{\rm bar}$ \textbf{(b)} distributions
 for the 13 barred S0 galaxies detected through ellipse fitting (including ones detected
  through relaxed criteria). The observed and deprojected values for the three bars
  detected through the ellipse fit relaxed criteria are shown as filled and open 
  circles, respectively ($\S$~\ref{barred}). The mean observed $a_{\rm bar}$ for our barred S0s (including 
those detected with relaxed criteria) is
  2.5$\pm$1~kpc (2.9$\pm$1~kpc deprojected), while the mean  observed and deprojected $e_{\rm bar}$ is 0.4$\pm$~0.1.  Most (85\%) of bars have an 
observed $e_{\rm bar}\le$~0.5. We also note that all extra bars that were detected via the relaxed ellipse fit criteria on the observed images, 
would be detected via the strict ellipse fit criteria after deprojection. This is due to the fact that the latter removes projection 
effects, which cause the maximum bar ellipticity $e_{\rm bar}$ to go from a local maximum in the radial profile of ellipticity 
to a global maximum. 
 \label{abarebar}}
\end{figure}

\clearpage
\setcounter{figure}{12}
\begin{figure}
\epsscale{0.7}
\plotone{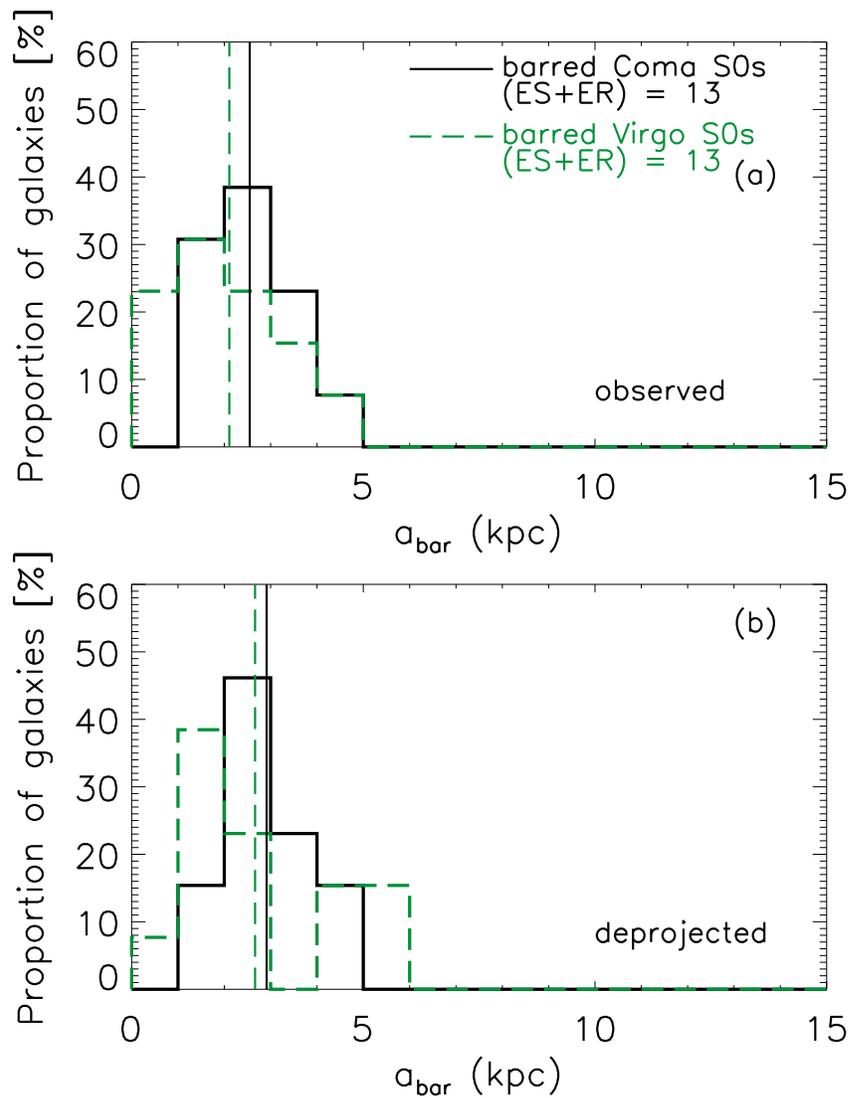}
\vskip 1 in 
\caption{Comparison of the observed \textbf{(a)} and 
deprojected \textbf{(b)} bar semi-major axis $a_{\rm bar}$
distributions for our sample of barred bright Coma core S0s and those in 
Virgo from Erwin et al.~(in prep.). For both samples, the distributions
include barred galaxies detected through the strict ellipse-fitting 
criteria as well as the relaxed criteria ($\S$~\ref{barred}). The 
vertical lines show the mean values for each distribution.  We do not find a significant
difference in the mean observed and deprojected bar size between Coma and Virgo S0s. The results from 
a KS test are consistent with similar distributions (giving KS ($p = 0.5, D=0.3$) and (0.2, 0.4) for panels (a)
and (b), respectively.
 \label{abarPEcomp}}
\end{figure}

\clearpage
\setcounter{figure}{13}
\begin{figure}
\epsscale{0.7}
\plotone{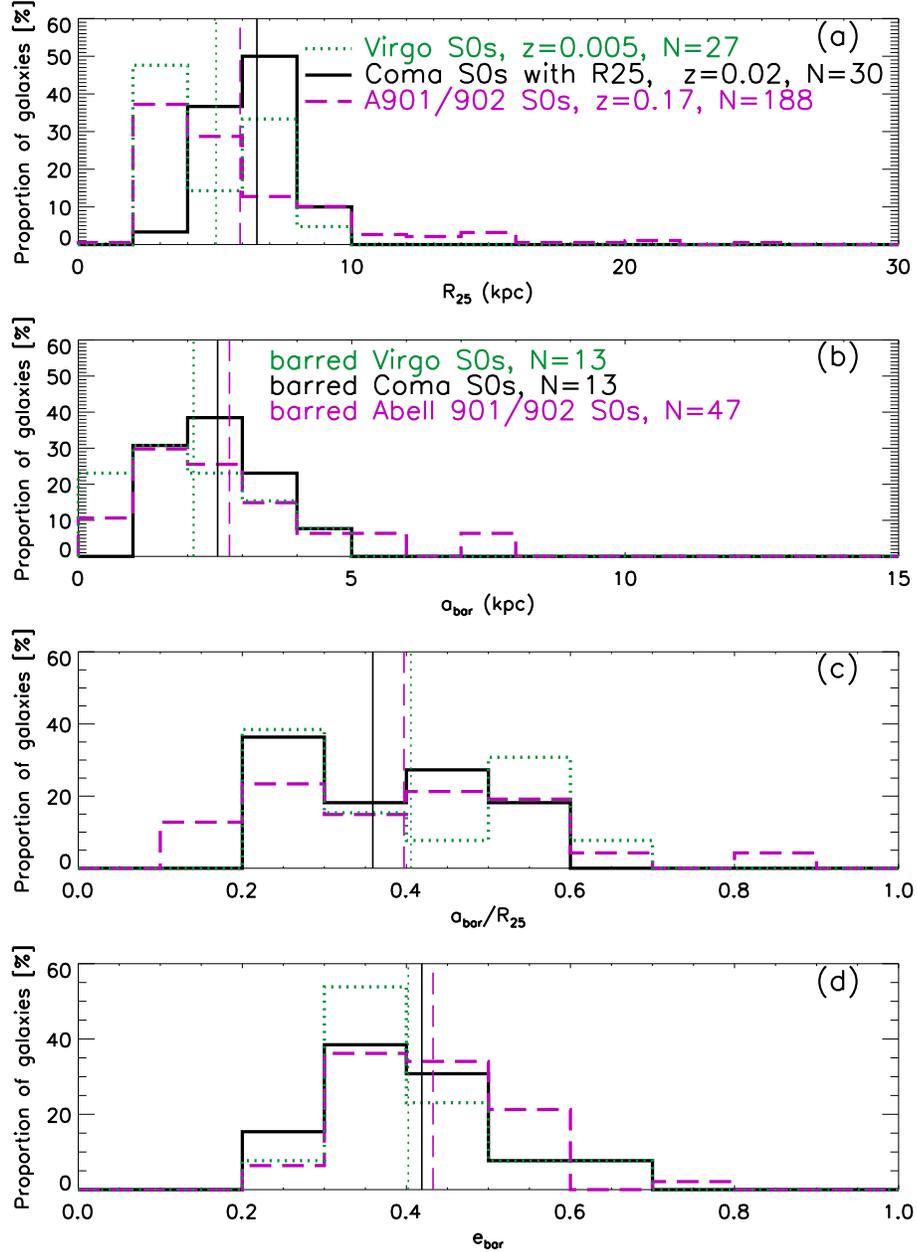}
\vskip 1 in 
\caption{Distributions of \textbf{(a)}~disk $R_{\rm 25}$ (the isophotal radius
where $\mu_{\rm B}$ reaches 25~mag~arcsec$^{-2}$), \textbf{(b)}~bar semi-major axis $a_{\rm bar}$, measured at the peak 
bar ellipticity $e_{\rm bar}$ for all bars identified 
through ellipse fitting (ER+ES), \textbf{(c)}~$a_{\rm bar}$/$R_{\rm 25}$ ratio, and 
\textbf{(d)}~peak bar ellipticity $e_{\rm bar}$ for 
the Coma S0 sample (solid black) and the comparison samples of S0s from the intermediate-density
cluster system Abell 901/902 (dashed pink) and the low-density Virgo cluster (dotted green). $R_{\rm 25}$ 
values for Coma and Abell 901/902 S0s are derived as described in $\S$~\ref{barredpropdisc}, while
$R_{\rm 25}$ for Virgo galaxies are from the RC3. The vertical lines
show the mean values for each distribution. All three samples have 
similar mean bar and disk properties, but the bar semi-major axis
and disk $R_{\rm 25}$ distributions for Abell 901/902 S0s 
have a tail to larger values. The KS statistic reflects the
differences in the Coma, STAGES, and Virgo distributions of $R_{\rm 25}$, giving ($p =
0.008$, $D=0.5$) between Coma and Virgo and ($p = 10^{-4}$, $D=0.4$) between Coma and
Abell 901/902. 
 \label{STVbardisk}}
\end{figure}

\clearpage
\setcounter{figure}{14}
\begin{figure}
\epsscale{0.8}
\plotone{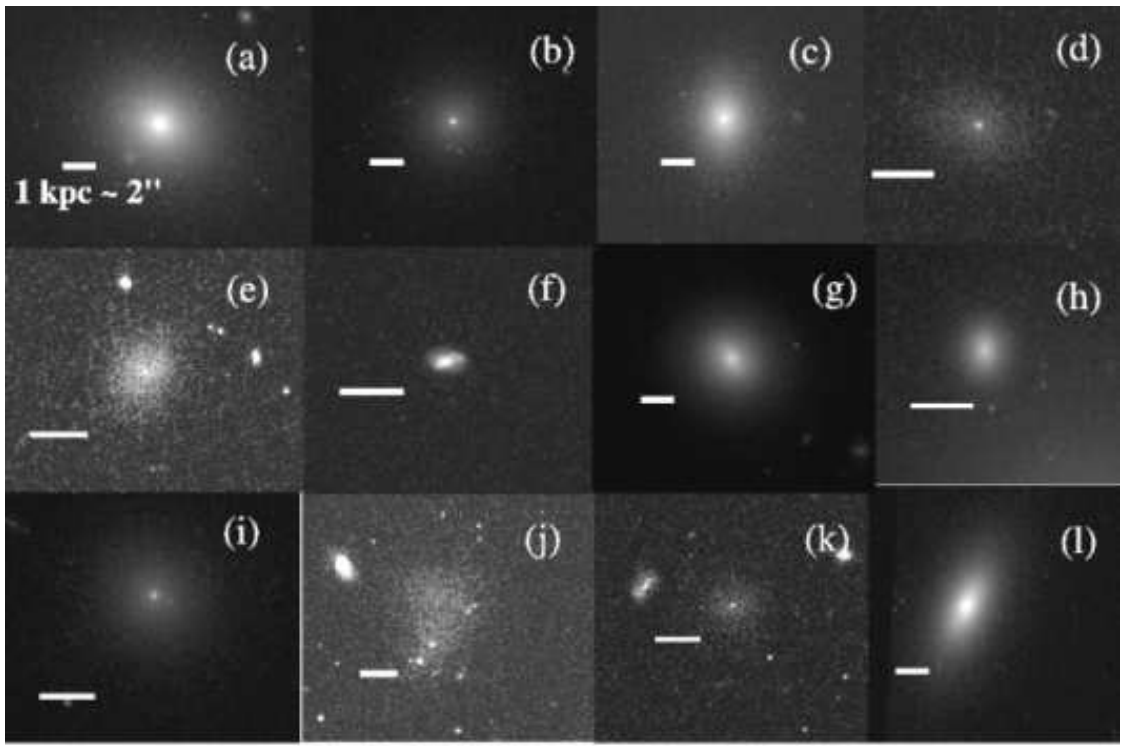}
\caption{Examples of some of the faint, low-mass dwarfs in the Coma core sample. 
The scale bars are 1~kpc. The galaxies shown are:
(a)~COMAi13011.143p28354.92, (b)~COMAi13025.977p28344.68,
(c)~COMAi13026.152p28032.02, (d)~COMAi13029.853p28400.85,
(e)~COMAi13030.027p28135.08, (f)~COMAi13039.068p28437.52,
(g)~COMAi13041.192p28242.38, (h)~COMAi13047.670p28533.95,
(i)~COMAi13048.045p28557.42, (j)~COMAi13050.590p28356.56,
(k)~COMAi13052.942p28435.86, (l)~COMAi13030.949p28630.18. 
 \label{dwarfmont}}
\end{figure}

\clearpage
\setcounter{figure}{15}
\begin{figure}
\epsscale{0.6}
\plotone{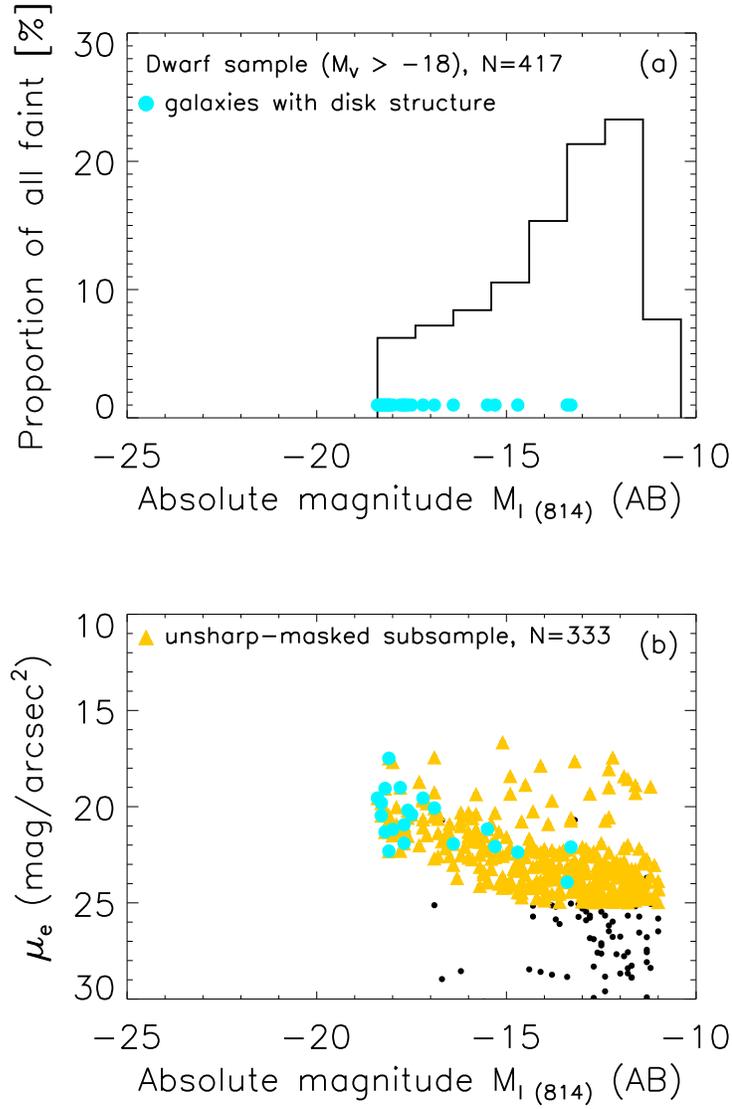}
\caption{\textbf{(a)}~Absolute magnitude ($M_{\rm I (814)}$)
  distribution and \textbf{(b)}~plot of surface brightness vs. absolute magnitude ($M_{\rm I (814)}$) of 
  the 417 galaxies in the Coma faint sample. The cyan points show the values
  for the 21 faint galaxies where we find disk structure (bar, spiral, edge-on disk)
  through unsharp masking. Most (76\%) of the objects where we find disk structure have  $M_{\rm I (814)} \le -16$.
 \label{dwarfmag}}
\end{figure}

\clearpage
\setcounter{figure}{16}
\begin{figure}
\epsscale{0.9}
\plotone{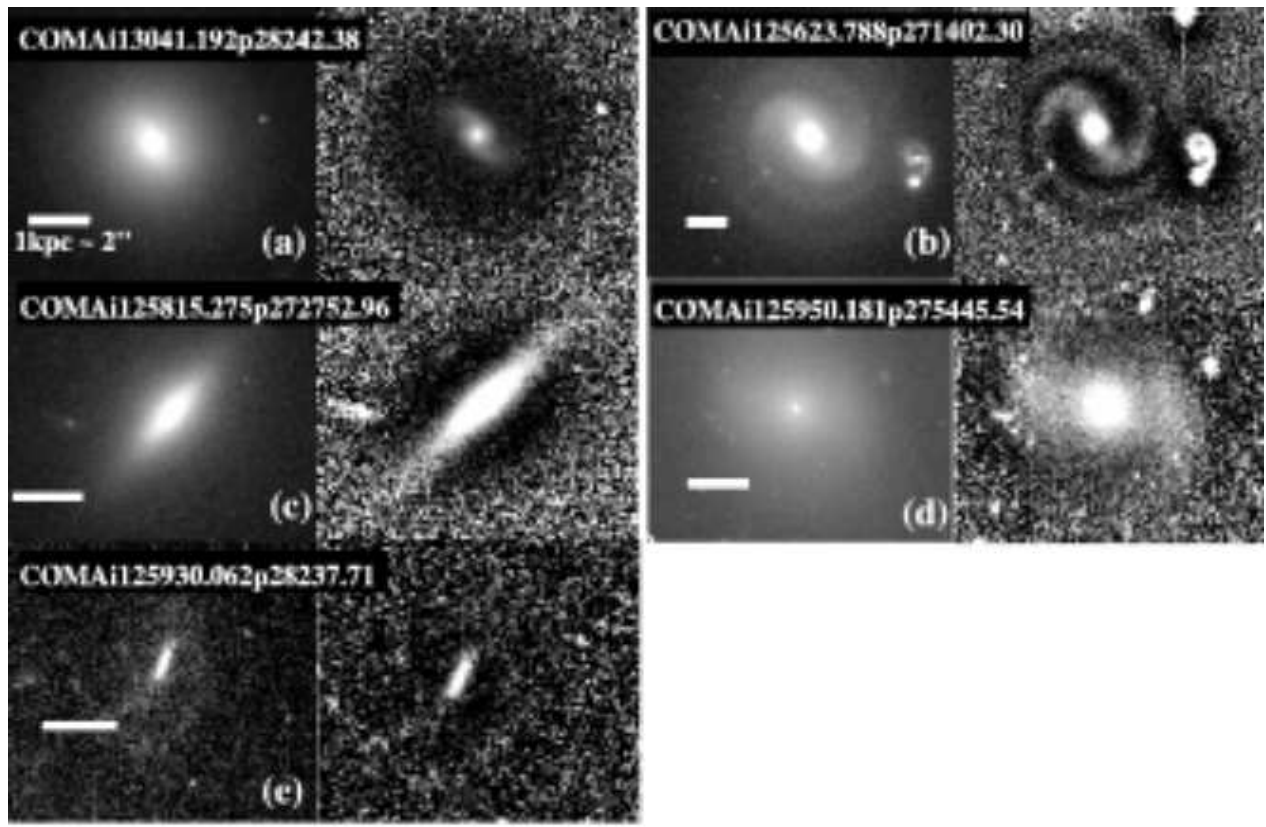}
\caption{Examples of five galaxies from the dwarf ($M_{\rm V} > -18$) 
sample highlighting the different types of 
disk structure that we find through unsharp masking: \textbf{(a)}~spiral arms, 
\textbf{(b)}~bar+spiral arms, 
\textbf{(c)}~edge-on disks, \textbf{(d)}~bar and/or spiral structure, and \textbf{(e)}~ambiguous bar/edge-on disk ($\S$~\ref{dwarfmethod}). 
We use a Gaussian smoothing kernel size of
$\sim$~25 pixels, corresponding to $\sim$~625~pc at the distance of
Coma. The original $HST$ images are shown in the left panels and the
corresponding residuals highlighting the disk structure are
on the right.  The scale bars show 1~kpc. 
 \label{usmask}}
\end{figure}

\end{document}